# GIFfluence:

# A Visual Approach to Investor Sentiment and the Stock Market*

Ming Gu
Hong Kong Polytechnic University

David Hirshleifer
Marshall School of Business, University of Southern California

Siew Hong Teoh
UCLA Anderson School of Management

Shijia Wu
Chinese University of Hong Kong, Shenzhen

December 21, 2025

We study dynamic visual representations as a proxy for investor sentiment about the stock market. Our sentiment index, GIFsentiment, is constructed from millions of posts in the Graphics Interchange Format (GIF) on a leading investment social media platform. GIFsentiment correlates with seasonal mood variations and the severity of COVID lockdowns. It is positively associated with contemporaneous market returns and negatively predicts returns for up to four weeks, even after controlling for other sentiment and attention measures. These effects are stronger among portfolios that are more susceptible to mispricing. GIFsentiment positively predicts trading volume, market volatility, and flows toward equity funds and away from debt funds. Our evidence suggests that GIFsentiment is a proxy for misperceptions that are later corrected.

---

* We thank Joey Engelberg, Jieying Hu, William Mullins, Litong Zhang and conference and seminar participants at the Institute of Financial Studies at Southwestern University of Finance and Economics; UCSD Rady School of Management; INSEAD; HEC Paris Business School; the *China Journal of Accounting Research* Annual Conference in Shenzhen; and the Hawaii Accounting Research Conference in Honolulu for very helpful comments.

# 1 Introduction

Social media users often use pictures or short videos to convey ideas and emotions to others. The dynamic visual representations provided by investors reflect the thoughts and feelings that underly trading decisions in the financial markets. This study examines the relation between dynamic visual representations in the social media communications of investors and stock market outcomes.

To develop a visual measure of investor sentiment, we focus on the Graphics Interchange Format (GIF), which uses short looping video animations to vividly convey emotions and ideas, often with a humorous twist.[1] GIFs are widely shared and highly effective at expressing sentiment (Milner and Highfield 2017). Our investor sentiment index, GIFsentiment, is based on the GIFs shared in social media communications about the stock market. We use this novel index to test whether sentiment predicts stock market returns, trading volume, volatility, and equity versus bond fund flows. To our knowledge, this is the first study to directly relate GIF visuals to market sentiment and stock market outcomes.

GIFs are powerful communication tools. In some ways they are more effective than text in conveying ideas and feelings. Short, emotionally laden animations tend to engage fast, automatic thinking (referred to as System 1 in the dual-process cognition framework of Kahneman 2011) rather than the effortful reasoning processes of System 2. GIFs integrate motion and sequential images to uniquely encapsulates ideas about past events, future forecasts, or cause-and-effect relationships in the form of mini-stories, making them highly engaging.[2] Furthermore, the salience of GIFs makes them especially effective in capturing the attention of receivers.

Neuroscience research on social cognition shows that visual stimuli have a profound effect on attention and thought (Fiske and Taylor 2020; Dragoi and Tsuchitani 2016). A large part of the human cerebral cortex is dedicated to processing visual stimuli. Tech industry leaders have also emphasized that combining visuals with text (multimodality) is key for effective communication and message delivery.[3]

---

[1] See https://gmis.me/Animated_GIF_Examples_and_their_Static_Counterparts.htm for some examples comparing GIFs and still images. An anecdotal example of the association of GIF use with market outcomes is the 6.3% drop in the stock price of Tesla on the day a GIF of Elon Musk apparently smoking marijuana in a Joe Rogan podcast went viral; see the GIF at https://giphy.com/gifs/weed-blunt-elon-musk-2Y8Iq3xe121Ba3hUAM. Figure 1 provides static frames from four illustrative GIF examples.

[2] Images and motion are triggers for bottom-up attention, which is an effortless and automatic response to a salient stimulus (see, e.g., Li and Camerer 2022). GIFs can also engage with top-down attention, which is effortful and deliberative, as GIFs can highlight a key idea or topic of interest to the recipient.

Furthermore, GIFs portray motion, which is a powerful attentional cue. Motion triggers physiological arousal, thereby increasing the likelihood of action or response. Motion elicits strong physiological arousal as measured by skin conductance (Detenber, Simons, and Bennett Jr. 1998; Fox et al. 2004; Simons et al. 1999). More generally, research on physiological arousal shows that elevated arousal is associated with more extreme evaluations, enhanced long-term memory (Storbeck and Clore 2008), increased risk-taking (FeldmanHall et al. 2016), autonomic responses and impulsive decision-making (Herman, Critchley, and Duka 2018). These findings suggest that the use of GIFs may induce impulsive investor decisions.

GIFs are vivid in the sense of being emotionally engaging, concrete, and imagery-rich sources of information (on vividness, see Nisbett and Ross 1980). Images tend to elicit emotional responses that are faster and stronger than those evoked by words, as documented in neuroscience studies, including EEG- and fMRI-based evidence (Kensinger and Schacter 2006; Schacht and Sommer 2009). Emotions affect optimism and risk tolerance, and thereby investor decisions (Lerner et al. 2015; Wachter and Kahana 2024). The succinctness (Potter et al. 2014) of GIFs and their distinctive immediacy of consumption (Bakhshi et al. 2016) enhance emotional intensity, making GIFs instantly understandable and highly engaging in expressing emotions and telling stories.

Finally, GIFs also can depict sequences of events, making them well suited to conveying understandings of causes of past or future events and simple mental models or narratives about the stock market (Shiller 2017; Hirshleifer 2020; Andre, Schirmer, and Wohlfart 2024). For example, a GIF of a rocket launching toward the moon can represent an anticipated rapid rise in stock price, thereby inducing or reflecting investor sentiment.

[3] See, for example, the discussion between OpenAI co-founder Ilya Sutskever and Nvidia Founder/CEO Jensen Huang in Highlights of the Fireside Chat AI Today & Vision of the Future starting at 00:23:47.

Based on these considerations, we use GIFs to construct a novel proxy for investor sentiment. We introduce a daily aggregate market-level investor sentiment index, GIFsentiment, derived from GIFs embedded in messages posted on Stocktwits.com, a leading online platform for sharing opinions about stocks and financial assets. We then examine the relation between GIFsentiment and market outcomes in comparison with other established sentiment proxies.

Behavioral theories predict that an investor sentiment proxy will be positively associated with contemporaneous returns as overvaluation grows and will negatively predict returns in subsequent periods when overvaluation is corrected. To evaluate whether GIF sentiment is an investor sentiment proxy, we therefore test whether GIF sentiment has a positive contemporaneous association with equity index returns and whether it negatively forecasts subsequent returns. We further test whether such relationships hold incrementally after controlling for five other sentiment proxies (see below and Section 2.3) and attention proxies from past literature. We also examine whether the association of GIF sentiment with returns differs among aggregate indices that differ in size or idiosyncratic volatility. Finally, we test whether GIF sentiment is related to contemporaneous and subsequent trading volume, market volatility, and equity and bond fund flows.

To classify GIF sentiment on a large scale, we exploit a feature on Stocktwits that allows users to self-declare their posts as bullish or bearish. We define a unique GIF as a specific animated image (e.g., a rocket shooting toward the moon), treated as a single unit regardless of how many times it appears across posts. For each unique GIF, we tally its appearances in bullish- and bearish-labeled posts and compute its net bullish sentiment. Crucially, this GIF-level net bullish sentiment allows us to gauge the net optimism of all posts that contain GIFs—even posts that do not have sentiment declarations. We then derive the aggregate market sentiment for that day, GIFsentiment, by combining the net optimism measures for all GIF-containing posts for any given date.

Specifically, we quantify each unique GIF's valence as the net-bullish share, defined as the difference between bullish and bearish declarations divided by the total appearances. To avoid look-ahead bias, we calculate this valence using a forward expanding window, updated daily. The aggregate daily GIFsentiment is the appearance-weighted average valence of these unique daily GIF sentiment scores; details in Subsection 2.2.

Our sample runs from September 2020, when Stocktwits introduced an in-composer GIPHY search button, through October 2024. GIPHY, one of the largest GIF repositories worldwide, allows users to search and insert GIFs easily, enabling more vivid expression of sentiment. Stocktwits also includes a dedicated bullish or bearish button throughout our sample period, allowing users to give a binary declaration of their sentiment as a part of their posts.

Our main tests assess the predictive power of GIFsentiment for market outcomes and its incremental contribution beyond the following five sentiment measures (see Section 2.3 for details): TEXTsentiment, monthly Baker and Wurgler (2006) sentiment measure BW, monthly University of Michigan consumer sentiment index ICS, MEDIAsentiment from RavenPack, and SELFDEC derived from self-declared bullish or bearish sentiment in postings without GIFs. These sentiment measures are pairwise correlated, suggesting that there is commonality in what they capture.[4]

The first set of tests describe the associations between the sentiment measures with several proxies for investor feelings identified in past research. The mood proxies include a calendar-month-based optimism indicator (Thaler 1987; Hirshleifer, Jiang, and DiGiovanni 2020), and pessimism indicators based on calendar months (Kamstra et al. 2017), cloud cover (Hirshleifer and Shumway 2003), and an index of change in the stringency index of government COVID lockdown restrictions (Terry, Parsons-Smith, and Terry 2020; Bueno-Notivol et al. 2021).

Consistent with GIFsentiment capturing investor mood, we find that GIFsentiment is lower on days with higher cloudiness and when government lockdown restrictions became stricter, and during months with declining mood. In contrast, the other sentiment measures do not exhibit consistent associations with the mood proxies, suggesting that these measures may not primarily

[4] Obaid and Pukthuanthong (2022) develop a proxy for investor sentiment based upon human ratings of still photos. As it is not straightforward to acquire large numbers of news media photos and generate human ratings of them, we do not use their sentiment measure as a control in our tests. In Section 2 we discuss their findings and some advantages of using GIF sentiment relative to hiring people to rate still images.

reflect variation in mood.

To test whether the six indexes reflect reactions to contemporaneous fundamental news as compared with pure sentiment, we estimate their associations with a measure of aggregate earnings news. We define a firm's earnings announcement outcome as non-negative if earnings meet or beat consensus analyst forecasts and define aggregate earnings news as the fraction of earnings announcement outcomes that are non-negative on the given day. We find that GIFsentiment and BW are not significantly correlated with aggregate earnings news, while the four other proxies are. These results suggest that, unlike GIFsentiment, most sentiment measures may reflect a mixture of investor mood/attention with reactions to fundamental news.

Our main tests examine the predictive power of sentiment for market outcomes motivated by the investor sentiment model of De Long et al. (1990). In this model, high investor sentiment produces market overpricing followed by low subsequent returns. This occurs because sentiment-driven investors increase their demand for risky assets, driving prices above fundamentals. An opposite dynamic occurs when sentiment is low.

Consistent with these predictions, we find that GIFsentiment is positively correlated with contemporaneous aggregate stock market returns and is a highly significant and strong negative predictor of market returns during the first month after the sentiment conditioning date. This negative return predictability suggests that our GIFsentiment measure captures mispricing rather than fundamental information. The magnitudes of the relationships are economically meaningful. A one standard deviation increase in GIFsentiment is associated with an additional 27.3 basis points on the contemporaneous S&P 500 index return, and a return that is lower by 126.5 basis points in the first month.[5]

There is no indication that economic fundamentals, prior investor attention, or other sentiment measures drive these relationships. We control for contemporaneous fundamental events using daily news-based measures of U.S. economic policy uncertainty, EPU (Baker, Bloom, and

[5] The fact that the post-conditioning date returns exceed the conditioning date returns suggests that high daily GIF sentiment may partly reflect sentiment and overpricing present prior to the conditioning date.

Davis 2016) and daily U.S. macroeconomic activity index, ADS (Aruoba, Diebold, and Scotti 2009). We also control for two proxies of investor attention: (1) attention to fundamental financial news, based on the number of earnings announcements on a given day (Hirshleifer, Lim, and Teoh 2009), and (2) general social media attention, based on number of messages posted relative to the preceding 10 trading days (Cookson, Lu, Mullins, and Niessner 2024). In addition, we control for the other five sentiment measures. Furthermore, similar findings hold in intraday specifications, in which GIFsentiment has a strong negative association with two-day forward returns.

In sharp contrast, the findings for the other sentiment measures are not fully consistent with the hypothesis that imperfectly rational investor sentiment induces temporary deviations from efficient pricing. For example, TEXTsentiment is positively associated with contemporaneous returns, yet there is no evidence of a subsequent one-month negative association with returns. BW exhibits the opposite pattern of a negative relation with contemporaneous, one-week and one-month returns.

Investor sentiment theory (De Long et al. 1990) further suggests that greater investor bias creates greater initial mispricing, and that limits to arbitrage limit the ability of rational investors to induce rapid correction toward fundamentals (Pontiff 1996; Shleifer and Vishny 1997). We therefore expect a stronger overreaction and correction dynamic amongst those assets most sensitive to investor psychological bias and those that are riskier and costlier to arbitrage. Small stocks are likely more sensitive to retail investor sentiment and less liquid, which limits arbitrage (Lee, Shleifer and Thaler 1991). High uncertainty stocks, as proxied by idiosyncratic volatility, are also likely more sensitive to investor sentiment shocks and riskier to arbitrage (Pontiff 1996; Baker and Wurgler 2006; Kumar 2009).

To test these hypotheses, we examine the return predictability of major US equity indices and portfolios constructed based on firm size, idiosyncratic volatility, and total return volatility. Consistent with the hypotheses, GIFsentiment exhibits stronger return predictive power for portfolios of stocks that are smaller or more volatile.

It has long been argued that bubble thinking by investor promotes stock return volatility.

For example, in the model of Barberis et al. (2018), as bubbles grow investors “waver”, resulting in eventual bubble collapse. Allowing for the possibility of either positive or negative bubbles, we hypothesize that when sentiment is extreme (high or low) returns become more volatile. Since mean GIFsentiment is normalized to zero, we test for the effect of extremity by estimating the relationship between absolute sentiment and subsequent realized market volatility. As an alternative hypothesis, if a sentiment proxy is intermixed substantially with information about fundamentals, extreme values of the measure could reflect high resolution of fundamental uncertainty. This would tend to reduce forward-looking volatility.

We find a positive relation between absolute GIFsentiment and return volatility over the week after the conditioning date. In contrast, we find a negative relation between absolute sentiment and next week’s stock market volatility for all the other sentiment measures except BW and MEDIAsentiment. These findings are consistent with GIFsentiment capturing investor sentiment rather than fundamentals, with the other sentiment measures, except for BW and MEDIAsentiment, reflecting at least in part fundamental information that resolves uncertainty.

We further hypothesize that more extreme sentiment promotes greater disagreement and trading activity, as sentiment-prone investors trade more heavily against sentiment-resistant investors. Consistent with this hypothesis, absolute GIFsentiment is positively associated with contemporaneous and subsequent trading volume.

Finally, we study the link between GIFsentiment and the behavior of retail investors as reflected in fund flows. Retail investors – often characterized as noise traders in past literature – participate heavily in fund investing, with 125 million holding mutual funds by year-end 2024 (Investment Company Institute 2025), and hold a large share of U.S. mutual fund assets. We hypothesize that increases in optimism, as proxied by GIFsentiment, predicts greater flows into equity relative to bond funds. Consistent with this hypothesis, we find that higher GIFsentiment predicts substantial subsequent-week inflows into equity funds and outflows from bond funds.

To sum up, we provide a new measure of aggregate market sentiment based on dynamic visual representations that predicts aggregate trading, fund flows, and stock market returns incremental to existing sentiment measures. Several authors have argued that variation in investor sentiment is driven in large part by social interaction (Shiller 2017; Hirshleifer 2020; Kuchler and Stroebel 2021; Cookson, Mullins, and Niessner 2024). As such, our paper contributes to the growing field of social finance. This field emphasizes that financial decisions and market dynamics are shaped by information and opinions transmitted through social interactions, which includes social media interactions (Hwang 2023) and the increasingly important role of visual media.

This paper builds on a literature that identifies predictors of aggregate stock market returns, such as dividend yield and investor sentiment proxies such as cloud cover, sports results, or media discourse.[6] Our paper differs in developing a dynamic visual-based sentiment measure from social media postings, GIFsentiment, which captures variations in mood and attention. This approach highlights the social transmission of investor sentiment, thereby reflecting how feelings spread from person to person rather than taking the sole form of isolated reactions to mass media.[7]

In contemporaneous work, Cookson et al. (2025) provide evidence that attention and sentiment measures derived from Twitter, Seeking Alpha, and StockTwits predict aggregate market returns. One way in which our paper differs is in examining visual content (GIFs), whereas their sentiment measure is a composite index based on textual analysis and bullish or bearish self-declarations. Their index negatively predicts returns in the next month, and their attention measure negatively predicts return on the next day. We find that GIF sentiment negatively predicts aggregate returns in the next month (even after controlling for both text-based and self-declaration-based sentiment). Cookson et al. (2025) also test how measures of attention versus sentiment relate to turnover. Our paper differs in testing how visually expressed sentiment relates to return volatility and bond and equity fund flows.

---

[6] Some papers use individual sentiment proxies to predict returns at horizons of up to one week (Hirshleifer and Shumway 2003, Edmans, Garcia, and Norli 2007, Da, Engelberg, and Gao 2015, and Obaid and Pukthuanthong 2022). Other individual proxies have been found to predict returns at horizons of many months, such as fund flows (Ben-Rephael, Kandel and Wohl 2012) and war discourse in the news media (Mai, Hirshleifer, and Pukthuanthong 2024). Huang et al. (2015) find that a combination of multiple sentiment proxies predicts one-month-ahead returns.

[7] Our paper also differs from some of these measures that are based in part on market prices. Market price based proxies are expected to predict returns as prices reflect both risk premia and expectations. As such, it is harder to clearly distinguish sentiment effects from risk premium effects using price-based measures.

As mentioned earlier, we also document that GIFsentiment in some ways captures investor feelings better than sentiment measures considered in past literature. GIFsentiment has consistent and significant correlations with established mood proxies, whereas the alternative sentiment measures do not. These findings align with the idea that GIFs offer a window into the affective dimension of investor communication that goes beyond what text alone conveys. This point is reinforced by the recent work of Gu et al. (2023), which uses firm level GIF-based sentiment measure to explain the cross section of individual stock expected returns. Our paper differs in testing the implications of aggregate GIFsentiment for aggregate trading and returns.

## 2 Sentiment Measures and Mood Proxies

This section details our sentiment measures and mood proxies, beginning with a comparison of GIFsentiment to existing literature (Subsection 2.1) and the construction of GIFsentiment (Subsection 2.2). We then describe alternative sentiment measures used as controls (Subsection 2.3) and evaluate GIFsentiment's validity through its association with established mood proxies (Subsection 2.4).

### 2.1 Other investor sentiment proxies

Previous research quantifies investor sentiment using surveys, combinations of economic variables, and textual analysis from various sources to forecast stock market returns and other outcomes. Other studies also employ event shocks, such as morning sunshine (Hirshleifer and Shumway 2003) and sports outcomes (Edmans, Garcia, and Norli 2007) as proxies for investor mood.

Survey-based measures include the Michigan Consumer Index (Qiu and Welch 2004) and the American Association of Individual Investors (AAII) monthly sentiment survey of the views of individual investors about the next six months. The classic Baker and Wurgler (2006) sentiment index, BW, is constructed from the first principal component of six economic variables– NYSE turnover, the dividend premium, IPO activity and their first-day returns, the closed-end fund discount, and the equity share in new issues. Sentiment has also been gauged from fund flows, as in the study of Ben-Rephael, Kandel and Wohl (2012).

Textual sentiment measures are constructed from media articles, financial reports, and social media postings, typically using word frequencies or expressed opinions (e.g., Tetlock 2007; Loughran and McDonald 2011; Da, Engelberg, and Gao 2015; Chen et al. 2014; Cookson and Niessner 2020). Sentiment has also been extracted from non-textual media, such as music billboard charts (Edmans et al. 2022).

A challenge in deriving sentiment measures from social media text or visuals is ecological validity: ensuring that social media context aligns with the investing context. General platforms like Twitter include a broad set of participants, many with limited or no interest or participation in the stock market. Similarly, Amazon Mechanical Turk (MTurk) raters are unlikely to be active investors, so the use of such raters to train machine-learning sentiment models creates a mismatch with the target investor population. Nevertheless, Obaid and Pukthuanthong (2022) find that a visual sentiment proxy derived from 148,823 static news photos from the *Wall Street Journal* (*WSJ*) predicts aggregate market returns one week ahead.[8] Their daily pessimism index was trained using 882 Twitter photos. This suggests an approach of dynamic visuals with improved context matching may capture sentiment more powerfully, and indeed our measure predicts returns over a much longer horizon.

Our method mitigates context mismatch by using Stocktwits' native environment, extracting GIF sentiment directly from users' self-declarations. These individual posters are typically active or interested stock market participants, ensuring ecological validity. The use of self-declarations also provides a natural 'ground truth' about investor attitudes that is unavailable for most other sentiment proxies in the literature. This approach also avoids concerns about the use of external raters such as undergraduates or MTurkers—high costs, small training sample size, limited incentives, and possible mismatch of the sample with equity investors owing to limited exposure to the stock market (Saravanos et al. 2021; Aguinis, Villamore, and Ramani 2021).

We test for the incremental predictive power of GIFsentiment for stock market outcomes controlling for other sentiment measures from the literature, including text-based sentiment, Baker-Wurgler sentiment, the Michigan Consumer Index, and traditional media sentiment. Their construction is described in Subsection 2.3. While other studies examine static visuals in corporate communications and their firm-level association with investor perceptions (Blankespoor, Hendricks, and Miller 2017; Nekrasov, Teoh, and Wu 2022; Peng et al. 2022; Christensen et al. 2024; and Ronen et al. 2023),[9] our study tests for a predictive relation between sentiment derived from dynamic visual representations and aggregate stock market outcomes.

[8] The authors validated the pessimism label assigned by their machine learning algorithm against labels assigned by five MTurk raters for a sample of 100 WSJ photos.

### 2.2 Construction of GIFsentiment

Launched in 2008, Stocktwits.com initially supported text and hyperlinks. In September 2020, users gained the ability to add GIFs via a menu button linking to Giphy.com, one of the largest global GIF search engines, enhancing their expressive capabilities. As shown in Figure 2, the proportion of postings containing GIFs rose over the initial three quarters and subsequently stabilized at approximately 10% of all postings through December 2024.

Stocktwits posts offer several features that facilitate the construction of a high-frequency sentiment measure. Each post has a date and time stamp, enabling daily sentiment measurement. Users can specify stocks via cashtags (e.g., $AAPL for Apple Inc.) and multiple cashtags can be used for discussions of several stocks. This feature helps identify stock-related posts, indicating broader attention to the equity market.

GIFs in Stocktwits posts are identifiable by their .gif URLs, and unique Giphy.com identifiers,[10] allowing for accurate tracking of individual GIFs. Our sample, from September 1, 2020 to October 31, 2024, comprises 65 million posts cashtags, of which 5.5 million contain visuals and 463,227 are unique. Stocktwits introduced a bullish or bearish declaration feature in September 2012. Building on other research that uses these declarations as a message sentiment proxy (e.g., Cookson and Niessner 2020, Cookson, Engelberg, and Mullins 2023, Cookson et al. 2024), we exploit posts with both GIFs and self-declarations to infer sentiment for the same GIFs in posts lacking explicit declarations.

---

[9] Blankespoor, Hendricks, and Miller (2017) analyze video clips of IPO roadshows to assess investor perceptions of CEOs. Nekrasov, Teoh, and Wu (2022) study the use by firms of static images in earnings-related tweets. Christensen et al. (2024) examine the types and placement of infographics in 10-K filings, and the factors influencing the use of inforgraphics over time. Ronen et al. (2023) introduce a machine learning-based measure to quantify the informativeness of images on equity crowdfunding pitch webpages, and to link image characteristics to fund investment decisions.

[10] Stocktwits partners with Giphy.com, the world's largest GIF search engine, to enable users to the select and post GIFs seamlessly. All GIF URLs on Stocktwits.com are hosted on Giphy.com and share a uniform hyperlink structure: https://media2.giphy.com/media/{gif.id}/giphy.gif.

We construct a daily sentiment label for each unique GIF using the daily count of associated bullish and bearish declarations, ensuring no look-ahead bias. Over our sample, 96% of GIFs had at least one declaration, 71% had five or more, and 50% had 25 or more. To ensure consensus, we include only GIFs with at least five declarations. This method essentially calibrates the positive or negative sentiment of each unique GIF by aggregating self-declared expressions of optimism and pessimism across multiple posts, enabling inference for GIF-containing posts without explicit declarations.

We calculate the sentiment of each unique GIF $j$ (Unique GIFsentiment$_j$) daily by subtracting bearish from bullish declarations and dividing by the total appearances for GIF $j$, using a forward-expanding window to prevent look-ahead bias. This provides a continuous daily net bullish sentiment measure for each unique GIF. Table A1 presents examples of GIFs with the highest and lowest sentiment scores. Aggregate daily GIFsentiment is then the appearance-weighted average valence for all GIF-containing posts on a given day:

$$\text{GIFsentiment}_t = \sum_j \left( \frac{\#\text{Appearance}_{jt}}{\#\text{GIFPosts}_t} * \text{Unique GIFsentiment}_{jt} \right), \quad (1)$$

where $\#\text{Appearance}_{jt}$ is the total number of posts for GIF $j$ on day $t$, $\#\text{GIFPosts}_t$ is the total number of GIF-containing posts on day $t$, and $\text{Unique GIFsentiment}_{jt}$ is the proportion of the net bullish sentiment declarations of GIF $j$ on day $t$. A trading day is defined with a 4:00 p.m. cutoff, aggregating posts from 4:00 p.m. on day $t$ - 1 to 4:00 p.m. on day $t$. Calculating daily GIF-level sentiment allows for temporal variation, accommodating shifts in market context, meme culture, or user interpretation.[11]

[11] Figure 3, Panel A presents the time series of daily GIF sentiment, SELFDEC, Text sentiment, and Media sentiment over our sample period from September 1, 2020 to October 31, 2024. In an untabulated analysis, we find that GIF

Temporal stability of GIF-level sentiment is assessed by examining the daily autocorrelation for a given GIF under varying minimum appearance thresholds. Table 1 Panel B shows that the average autocorrelation is 0.59 for GIFs with at least five appearances per day, rising to 0.79 for those with 25 or more. This demonstrates high temporal stability in GIF-level sentiment, particularly for frequently used GIFs.[12]

We hypothesize that GIFsentiment captures a unique dimension of investor behavior, distinct from sentiment conveyed through explicit textual declarations in non-GIF messages. Given the range of expressive formats chosen by Stocktwits users (GIFs only, bullish/bearish declarations only, both, or neither), we propose that users, as well as the posting contexts characterized by vivid, salient GIFs, exhibit a greater susceptibility to behavioral biases than those relying solely on textual sentiment declarations.

Supporting this distinction, Table A3 Panel A reveals that users who self-identify as novice investors on Stocktwits are more likely to use GIFs to express sentiment, while Panel B indicates that relative to novice investors, intermediate and professional investors favor explicit bullish or bearish declarations. This difference suggests that our GIF sentiment measure may be especially sensitive to the attitudes of those investors who are more prone to behavioral biases. To isolate behavioral effects, we also control for and compare GIFsentiment with SELFDEC, a declaration-based sentiment measure calculated exclusively from non-GIF postings.

## 2.3 Alternative sentiment measures

To evaluate GIFsentiment's incremental association with stock market outcomes, we control for various established sentiment measures. These include the Baker-Wurgler sentiment index (BW), the University of Michigan Consumer Sentiment Index (ICS), traditional media

---

sentiment does not exhibit any day-of-the-week seasonality. Panel B displays the time series of monthly GIF sentiment, BW sentiment, and ICS, which reveal some common trends across measures but also features specific to each sentiment indicator. Table 2, Panel B gives the correlations between GIF sentiment and the other five sentiment measures.

[12] In the market-outcome tests beginning with Table 4, we address possible serial correlation in residuals. Further details are provided in the discussion accompanying Table 4. Tables in the Online Appendix report randomized *p*-values following Nelson and Kim (1993) to address autocorrelation in GIFsentiment.

sentiment (MEDIAsentiment), textual sentiment from Stocktwits posts (TEXTsentiment), and self-declared sentiment from non-GIF Stocktwits posts (SELFDEC).

The BW index, a monthly broad-based measure of speculative sentiment, is derived from a principal components analysis of six market-wide indicators described earlier. ICS is a monthly survey measure of consumer confidence about the economy and financial conditions, available from the University of Michigan's Surveys of Consumers or from the Federal Reserve Bank of St. Louis FRED (Federal Reserve Economic Data) website. Both BW and ICS are from the preceding month relative to the GIFsentiment score.

Drawing from the text-based sentiment literature,[13] we include MEDIAsentiment, a daily measure of traditional news media tone provided by RavenPack. RavenPack uses a proprietary machine learning model and AI to code sentiment from words in news articles (as used by Jeon, McCurdy, and Zhao 2022; Bushman and Pinto (2024)). We also control for two Stocktwits-derived text-related sentiment measures. The first, TEXTsentiment, is a daily aggregate textual sentiment measure calculated as the daily average of VADER scores applied to Stocktwits posts.[14] The second is SELFDEC, a daily measure of self-declared bullish or bearish sentiment from Stocktwits posts excluding GIFs. SELFDEC is defined as the net bullish declarations from *non*-GIF posts, divided by the total number of such posts daily:

$$\text{SELFDEC}_t = \frac{\text{Bullish NonGIF Posts}_t - \text{Bearish NonGIF Posts}_t}{\text{NonGIF Posts}_t}. \tag{2}$$

The study of Cookson and Niessner (2020) on investor disagreement and the study of Cookson, Engelberg, and Mullins (2023) on echo chambers use a similar self-declaration measure without regard to the presence or absence of GIFs and find that self-declarations predict individual stock

[13] Sentiment measures have used the text of a *Wall Street Journal* column (Tetlock 2007), text from financial reports (Loughran and McDonald 2011), Google search terms (Da, Engelberg and Gao 2015), text from the long-form social media platform Seeking Alpha (Chen et al. 2014), and text from short-form social media platforms (Renault 2017; Giannini, Irvine, and Shu 2018).

[14] Valence Aware Dictionary and Sentiment Reasoner (VADER) is a lexicon and rule-based sentiment analysis tool. It uses a pre-defined sentiment lexicon containing over 7,500 words, phrases, and emoticons. Each word is assigned a valence score reflecting its positive, negative, or neutral sentiment intensity. VADER then calculates the average sentiment score for a given text body, which in our analysis is the text words in postings. Hutto and Gilbert (2014) find that VADER performs better than other tools in the setting of microblog content on social media. Sohangir, Petty, and Wang (2018) apply VADER to StockTwits and find that it outperforms SentiWordNet and TextBlob in classifying bullish and bearish sentiment.

returns or trading volumes.

All six sentiment measures are standardized to have zero mean and unit variance for comparability. Table 2 Panel A presents their distribution prior to standardization, and Panel B reports their pairwise correlations. The measures exhibit moderate correlations, ranging from 0.16 to 0.49, suggesting shared underlying components, yet distinct aspects of investor sentiment. GIFsentiment correlates most strongly with TEXTsentiment, BW, and SELFDEC.

## 2.4 Evaluating sentiment measures with mood proxies and fundamentals

Growing research highlights the influence of mood on investor behavior, with several studies proposing empirical proxies for investor mood. We consider daily-level and monthly-level mood proxies. Daily mood is affected by environmental factors such as cloud cover. Hirshleifer and Shumway (2003) find that cloudy weather correlates with lower aggregate stock returns in tests across 26 countries, supporting a mood-based interpretation. Accordingly, we construct a daily cloud cover measure by averaging hourly NOAA[15] sky cloud cover data between 6 a.m. and 12 p.m. across national weather stations (Goetzmann et al. 2015; Edmans et al. 2022). To isolate mood-relevant fluctuations from other seasonally varying effects, we de-seasonalize the cloud cover series by subtracting each week's average cloudiness from its corresponding daily values following Hirshleifer and Shumway (2003).

We also include a daily COVID-19 stringency index to capture mood variation from policy-induced psychological stress. Past research shows that pandemic restrictions adversely impacted population mood (Terry, Parsons-Smith, and Terry 2020; Bueno-Notivol et al. 2021; Edmans et al. 2022). We construct this index from lockdown restrictions compiled by the University of Oxford's COVID-19 government response tracker (available from https://github.com/OxCGRT/covid-policy-tracker/tree/master/data).

At the monthly level, we incorporate calendar seasonality in mood, drawing on psychological and behavioral finance literature. For the United States, January and March are associated with optimism and SAD (seasonal affective disorder) recovery, respectively, while September and October are associated with SAD onset and diminished mood. Accordingly, we define a positive mood indicator for January and March, and a negative mood indicator for September and October (e.g., Thaler 1987 on positive mood in January; Kamstra et al. 2017 and Hirshleifer, Jiang, and DiGiovanni 2020 on positive mood in January and March and negative mood in September and October).

[15] The National Oceanic and Atmospheric Administration (NOAA) provides local climatological data from over 1,000 weather stations. Each weather station records the degree of cloud cover, which takes on integer values of 0 (clear – no coverage), 1 (few – 2/8 or less coverage), 2 (scattered – 3/8-4/8 coverage), 3 (broken – 5/8-7/8 coverage), or 4 (overcast sky – 8/8 coverage).

To evaluate GIFsentiment as a mood proxy, we examine its association with established daily and monthly mood proxies. We compute correlations between GIFsentiment and weather-induced mood shifts, COVID-related restriction stringency, and seasons associated with mood variations. Parallel analyses using TEXTsentiment, SELFDEC, BW, ICS, and MEDIAsentiment benchmark the strength and uniqueness of the GIF-based construct.

Table 3 Panels A and B present the Pearson correlation estimates. Consistent with GIFsentiment capturing investor mood, Panel A Column 1 shows that GIFsentiment is negatively correlated with daily cloud cover (DCC) and COVID-related restriction stringency ($\Delta$COVID Index). Panel B Column 1 further reveals a negative association with Negative Months (September and October). For comparison, Columns 2 to 6 in Table 3 Panels A and B present correlations for the five alternative sentiment measures with the same mood proxies. (Since BW and ICS are monthly measures, their correlations are estimated at the monthly level using aggregated mood proxies.) These alternatives generally lack consistent, expected associations with mood proxies. Specifically, TEXTsentiment is negatively correlated with uplifted mood periods, contrary to expectations, and uncorrelated with DCC or $\Delta$COVID Index. SELFDEC and ICS are negatively correlated with the $\Delta$COVID Index, but not Positive Months, Negative Months, or DCC. While BW correlates with all four mood proxies in the predicted directions, none of the correlations are statistically significant. MEDIAsentiment has a positive correlation with DCC, contrary to expectations.

The inconsistent associations of these five established sentiment measures with mood variables are puzzling. This raises the question of what these sentiment measures actually capture. Some past studies suggest that textual sentiment measures may reflect fundamental news rather than purely reflecting investor mood (e.g., Tetlock 2007; Loughran and McDonald 2011).

To examine this possibility, we examine the relation between all six sentiment measures with economic fundamentals. While not strictly required, a low correlation with fundamentals suggests a purer proxy for sentiment-driven investor attitudes, as distinct from rational reactions to news. We use %PositiveEANews as a fundamental news proxy, defined as the percentage of earnings announcements that meet or exceed consensus analyst forecasts for cashtag-mentioned firms. Table 3 Panel C indicates that SELFDEC, ICS, and MEDIAsentiment are positively associated with %PositiveEANews at the 1% or 5% level, suggesting that they capture fundamental information rather than just behavioral sentiment, whereas TEXTsentiment (surprisingly) exhibits a negative association. In contrast, GIFsentiment and BW show no significant correlation with this fundamentals proxy. Taken together, the evidence in Panels A through C indicates that GIFsentiment, a construct derived from user-generated visuals, more purely reflects mood-driven sentiment as contrasted with rational responses to fundamental news.

## 2.5 Evaluating GIF sentiment using a GPT vision model

We further cross-validate GIFsentiment as a sentiment proxy by estimating its correlation with an alternative visual sentiment measure constructed by coding the valence of GIF sentiment using OpenAI's GPT-4o vision-language model, which was introduced after May 2024. Owing to the computational costs, we restrict this exercise to the top 1% of GIFs ranked by total appearance count. Our focus on the top 1% enhances representativeness by prioritizing frequently used GIFs over those appearing sporadically. We do the ranking of GIFs in the period prior to May 2024 to avoid introducing a spurious correlation between the two sentiment proxies.[16]

[16] After May 2024, the release of GPT-4o, which can generate or modify GIFs, could have affected the popularity of different GIFs. After GPT-4o became available, GIFs generated by users with GPT-4o could enter the sample and might artificially have a stronger association with self-declarations, as AI-generated GIFs are often created using explicit prompts (e.g., "create a bullish GIF") and are likely accompanied by bullish self-declarations in StockTwits

For this subset, we prompt GPT-4o to assign a continuous sentiment score ranging from −1 (strongly negative) to +1 (strongly positive) for each unique GIF. Each GIF receives a single score reflecting its overall emotional valence. We then calculate the correlation between this GPT-coded sentiment (GPT4o-GIFsentiment) and our self-declaration-based per-GIF sentiment measure (GIF-Level sentiment), defined as the average sentiment for each GIF across the estimation window. In addition to the continuous measure, we also construct binary sentiment indicators (−1 for negative sentiment score < 0, and +1 for positive sentiment score > 0) for both measures.

Table A3, Panel C, reports a correlation of 0.306 ($p < 0.01$) correlation between continuous GIFsentiment and GPT4o-GIFsentiment, with the binary indicators showing an even stronger correlation of 0.537 ($p < 0.01$). These findings indicate a high consistency of our GIFsentiment measure with a proxy for sentiment based upon an advanced vision-language model.

## 3 GIF Sentiment and Stock Returns

### 3.1 GIF sentiment association with stock market returns

This section examines the relation between GIFsentiment with contemporaneous daily returns on the aggregate CRSP value-weighted S&P 500 market index, SPX, and the ability of GIFsentiment to forecast future daily returns over the subsequent week and month. We analyze GIFsentiment individually and alongside five other sentiment measures to assess its incremental explanatory power. Investor sentiment theory posits that net positive sentiment is associated with positive short-term returns but predicts negative long-term underperformance as mispricing corrects. We run the following regression:

$$\%\text{Ret}_{(t+m,t+n)} = \alpha + \beta\ \text{Sentiment}_t + \gamma\ \text{Controls}_t + \varepsilon_t, \quad (3)$$

where $\%\text{Ret}_{(t+m,t+n)}$ represents contemporaneous day $t$, one-week ($t + 1$, $t + 5$), and one month ($t + 1$, $t + 20$) cumulative SPX returns. $\text{Sentiment}_t$ is $\text{GIFsentiment}_t$ in the individual regressions, posts. This could mechanically inflate the correlation between GPT-4o–coded sentiment and our original GIFsentiment.

or a vector including also TEXTsentiment, SELFDEC, BW, ICS, and MEDIAsentiment in joint regressions (β vector in that case). All sentiment measures are standardized to have zero mean and unit variance for comparability. The Variance Inflation Factor (VIF) values, ranging from 1.5 to 2, indicate that multicollinearity is not a concern. Behavioral theory predicts a positive coefficient for contemporaneous returns and negative coefficients for subsequent weekly and monthly returns.

Following past studies (Da, Engelberg, and Gao 2015; Edmans et al. 2022), we control for EPU, a daily news-based measure of U.S. economic policy uncertainty (Baker, Bloom, and Davis 2016),[17] ADS, a U.S. macroeconomic activity index from the Federal Reserve website (Aruoba, Diebold, and Scotti 2009),[18] past returns, %Ret[-5, -1] and %Ret[-21, -6], to account for potential return reversals, daily Log#EA, the logarithm of the number of earnings announcements, for investor distraction by fundamental events (Hirshleifer and Teoh 2003), and the daily abnormal number of posts on Stocktwits (Cookson et al. 2024) for general social media attention.

For the weekly and monthly returns predictability tests, these dependent variables have overlapping windows, which may induce serial correlation in the residuals. To address this issue, we estimate the return regressions using the moving block bootstrap method (Politis and Romano 1994) and report these tests in all tables beginning with Table 4. We choose the block size to match the length of the return horizon (five trading days for weekly returns and twenty trading days for monthly returns), ensuring that each bootstrap block preserves the autocorrelation structure generated by the overlapping return windows. To address potential serial correlation in the sentiment variables, we apply the Nelson and Kim (1993) randomized *p*-value procedure to the regressions in Table 4 and all subsequent tables. The randomized *p*-values are reported in the online appendix.

[17] This measure is constructed by counting the number of U.S. newspaper articles achieved by the NewsBank Access World News database with at least one term from each of the following three categories: (i) "economic" or "economy"; (ii) "uncertain" or "uncertainty"; and (iii) "legislation," "deficit," "regulation," "congress," "Federal Reserve," or "White House." Baker, Bloom, and Davis (2016) provide evidence that EPU captures perceived economic policy uncertainty. The data are available at https://www.policyuncertainty.com/index.html.

[18] This index extracts the latent state of macroeconomic activity from many macroeconomic variables (jobless claims, payroll employment, industrial production, personal income less transfer payments, manufacturing and trade sales, and quarterly real gross domestic product) using a dynamic factor model. The data are available at https://www.philadelphiafed.org/surveys-and-data/real-time-data-research/ads.

Table 4 Panel A reports results for the separate regressions of the returns on the Standard and Poor 500 Index (SPX) returns contemporaneously, one week following, and one month following the conditioning date on GIFsentiment and controls. GIFsentiment is associated positively with contemporaneous aggregate market returns and negatively with the subsequent one-month returns after controlling for fundamentals EPU, ADS, lagged returns, Log#EA, and Log#AbnMessages. The negative predictive power of GIFsentiment for future returns is consistent with GIFsentiment capturing mispricing rather than fundamental information. The estimated coefficients on the control variables are reasonable or as expected.

The economic magnitudes of the GIFsentiment coefficients are substantial. A one standard deviation increase in GIFsentiment is associated with a 27.3 basis points (bp) increase in same-day market returns, followed by a 126.5 bp lower return in the following month. This monthly reversal annualizes to −16.3%, a substantial economic magnitude.

The magnitude of the subsequent correction exceeds the contemporaneous reaction. This is plausible in a behavioral context where high sentiment at a given date is associated with preexisting overpricing as well as a new increment to mispricing at the focal date. Untabulated results further indicate that GIFsentiment positively correlates with prior week returns, consistent with sentiment persistence and associated mispricing. As mentioned earlier, we address serial correlation in the sentiment measures by computing robust *p*-values using the randomization procedure of Nelson and Kim (1993), and report these results for all regressions starting from Table 4 in the online appendix. The findings remain robust when using randomized *p*-values.

Panel B confirms that the relationship of returns with GIFsentiment is robust even after controlling for other sentiment proxies. A one-standard-deviation higher GIFsentiment is associated with a 29.1 bp higher same-day return ($p = 0.02$) and a 117.2 bp lower return over the subsequent month ($p = 0.05$). This sustained explanatory and forecasting power indicates that GIFsentiment is a strong incremental proxy for identifying past market overvaluation that corrects within a month.

Unlike GIFsentiment, other sentiment proxies do not display incremental return patterns consistent with investor sentiment theory. For instance, TEXTsentiment and MEDIAsentiment positively associated with contemporaneous (day 0) returns, but they do not significantly negatively predict future one-month returns, as expected under mispricing correction. BW and ICS are negatively associated with day 0 returns, contrary to sentiment theories, potentially owing to low power for a monthly sentiment measure. Overall, these findings indicate that GIFsentiment aligns more strongly with behavioral predictions, serving as a more effective proxy for investor sentiment.

### 3.1.1 Positive vs negative GIF sentiment

While many past sentiment measures focus on pessimism and negative market outcomes (Tetlock 2007; Chen et al. 2014; Da, Engelberg and Gao 2015; Obaid and Pukthuanthong 2022), high-arousal positive content is more likely to go viral (Berger and Milkman 2012). Social media users are more likely to upvote or retransmit positive messages (Kramer, Guillory, and Hancock 2014; Rosenbusch, Evans, and Zeelenberg 2019; Goldenberg and Gross 2020).

To separately explore the ability of GIFs to capture optimism versus pessimism, we construct separate positive and negative aggregate GIFsentiment measures from unique GIFs. Here, PositiveGIFsentiment is constructed as in Equation (1) using only GIFs with positive sentiment, and NegativeGIFsentiment is constructed using only GIFs with negative sentiment. A higher coefficient on PositiveGIFsentiment relative to NegativeGIFsentiment would indicate that one association is asymmetrically stronger than the other.

Table 5 presents the predictability of returns from PositiveGIFsentiment from net positive GIFs and NegativeGIFsentiment from net negative GIFs. Panel A shows both proxies predict returns in the predicted directions when included in regression Equation (3).

Notably, higher NegativeGIFsentiment (less intensely negative sentiment) predicts lower returns within one week, as shown in column (2), while higher PositiveGIFsentiment predicts lower returns over a longer one-month horizon in column (3). The reason for this asymmetry is an interesting topic for future research.

Panel B shows that with all six sentiment measures included in the regressions, the coefficient on PositiveGIFsentiment remains consistent with our main finding. In contrast, NegativeGIFsentiment is positively associated with day 0 returns but does not exhibit future return reversals. This suggests that its mood component may overlap or be subsumed by other sentiment measures included in the regression specification.

### 3.1.2 Intraday analysis

A caveat to our daily stock return analysis is the relatively short sample period, from September 1, 2020 to the end of 2024. This limits sample size, reducing statistical power. To partially address this, we investigate intraday GIFsentiment variation. For the intraday analysis, GIFsentiment, TEXTsentiment, and SELFDEC are constructed at 30-minute intervals. BW, ICS, and MEDIAsentiment, however, remain at their daily or monthly frequencies owing to data constraints. We examine the association of intraday sentiment with the SPY ETF returns from the TAQ database over three windows: (i) contemporaneous 30-minute interval (ii) short-term forward ($t$ + 30 minutes to $t$ + 1 day) and (iii) short-term forward ($t$ + 30 minutes to $t$ + 2 days). As noted earlier, the overlapping return windows can induce residual serial correlation, which we address using the moving block bootstrap (Politis and Romano 1994).

Table 6 presents the results. Panel A indicates that 30-minute GIFsentiment is positively associated with contemporaneous returns ($p < 0.01$) and higher GIFsentiment predicts negative returns over the subsequent two days. Panel B, with additional sentiment controls, indicates GIFsentiment continues to negatively predict returns over the following two days. These findings are consistent with our earlier analysis using daily GIFsentiment, supporting the robustness of main results against sample size or estimation instability concerns.

### 3.1.3 Robustness tests

We perform several robustness checks to ensure our main inferences are not influenced by outliers in returns or GIFsentiment, or sensitive to the market index. These include: (i) to mitigate concerns about anomalous early adoption patterns or unusual behavior during the 2020 meme stock period, we exclude the initial three months GIF adoption period (2020Q3 to 2020Q4) on Stocktwits when GIF usage likely took time to gain traction, results remain consistent (Table A7). (ii) We winsorize returns at the top and bottom 1%, 5%, and 10% levels; GIFsentiment remaining robust (Table A4). (iii) We construct GIFsentiment using only GIFs exceeding the 50th or 75th percentile of cumulative appearances, also yielding consistent results (Table A5). (iv) We excluding days with high DFBETA influence (absolute DFBETA > 2/ √ *n*,); GIFsentiment coefficients yield similar results (Table A6).

Overall, our findings are robust to outliers and early adoption platform dynamics or influence of the meme stock episode of 2020. Furthermore, untabulated results confirm that daily GIFsentiment-return relations are robust across all five stock indices: the CRSP value-weighted index (VWRETD), the SPDR S&P 500 (SPY), the PowerShares QQQ Trust (QQQ) for the portfolio of innovation stocks, the Russell 100 Index ETF (IWB), and the Russell 2000 Index ETF (IWM). This shows that the predictive power of GIFsentiment holds across indices with varying size and sector compositions.

### 3.2 GIF sentiment and susceptibility to mispricing

Investor sentiment theory predicts that the return forecasting power of a sentiment proxy is stronger for portfolios of stocks with greater mispricing pressure or tighter limits to arbitrage. Specifically, we test examine the association of GIFsentiment with immediate and long-term returns across stock portfolios that differ in firm size and idiosyncratic uncertainty. High uncertainty amplifies individual investor biases (Kumar 2009), while tighter limits to arbitrage increase mispricing (Pontiff 1996; Shleifer and Vishny 1997). Small stocks, being riskier, more costly to arbitrage, and disproportionately held by retail investors, are particularly prone to mispricing (Lee, Shleifer and Thaler 1991).

Motivated by these insights, we test whether the sensitivity of relationship between contemporaneous and future returns of stock index portfolios with GIFsentiment differs across portfolios that differ in these characteristics. To do so, we first stocks into market capitalization quintiles. Consistent with past research showing stronger sentiment-return associations for smaller stocks (e.g., Baker and Wurgler 2006, Edmans, Garcia, and Norli 2007), we expect stronger GIFsentiment return reversals in the small-stock quintile.

Next, we sort stocks into quintiles based on two risk measures: idiosyncratic volatility from Fama-French five-factor model residuals over the prior 36-months and total return volatility (Wurgler and Zhuravskaya 2002). We hypothesize a stronger association between GIFsentiment and returns among portfolios with higher idiosyncratic or total return volatility (Wurgler and Zhuravskaya 2002, Baker and Wurgler 2006).

To test these hypotheses, we estimate Equation (3) for subsamples based on firm size, idiosyncratic volatility, and total return volatility. We consider (1) the Small Cap (bottom quintile) and Large Cap (top quintile) groups; (2) High and Low idiosyncratic risk groups (top and bottom quintiles) and (3) HighVol and LowVol groups (top and bottom total return volatility quintiles). Table 7 reports these results.

Table 7 Panel A1 presents results for Small Cap (Columns 1-3) and Large Cap (Columns 4-6). GIFsentiment is positively associated with same-day returns and negatively predicts subsequent one-month returns for both groups, consistent with behavioral bias. Furthermore, Small Cap stocks exhibit larger magnitudes for both the contemporaneous positive returns and the negative one-month reversals. The last row in Panel A1 reports the *t*-statistics for the differences in coefficients between large and small caps across different return windows. The difference is statistically significant for the contemporaneous return window but not for the subsequent one-month window, although the magnitude is larger for small-cap firms.

Panel A2 indicates that the predictive association between GIFsentiment and returns remains economically meaningful after controlling for alternative sentiment proxies. A one-standard-deviation greater GIFsentiment is associated with a 18.8 bp higher day 0 return and a 275.6 bp lower return over the subsequent month for the small-cap group relative to the large-cap group. These differences are economically sizable, but owing to lack of power they are not statistically significant.

Panels B1 and B2 compare the associations of GIFsentiment with returns for high (Columns 1-3) and low (Columns 4-6) idiosyncratic volatility stocks. Estimates indicate a more pronounced association between GIFsentiment and both same-day and one-month-ahead returns for high-volatility stocks. The coefficient differences between the two groups are statistically significant, as indicated by their *t* statistics.

Panels C1 and C2 show similar patterns for firms grouped by total return volatility. Specifically, high-volatility stocks (top quintile) exhibit statistically and economically significant higher same-day returns and lower returns over the subsequent month compared to low-volatility stocks (bottom quintile).

### 3.3 GIF sentiment and stock market volatility

It has long been argued that investor behavior during bubble episodes can heighten return volatility. In the framework of Barberis et al. (2018), for example, as prices rise far above fundamentals, investors place shifting weight on recent price trends versus signals of overvaluation, creating instability that can culminate in a sharp reversal. Motivated by the idea that bubbles induce volatility, we consider both unusually high and unusually low sentiment as departures from typical belief levels. We hypothesize that more extreme GIFsentiment, whether positive or negative, is associated with greater subsequent volatility. Return volatility (%) is measured as the standard deviation of daily S&P 500 index returns over windows: from day $t$ through $t$ + 5, and from day $t$ through $t$ + 20, aligning with the windows in our return tests. We estimate the following regression:

$$\%\text{Volatility}_{[t,t+n]} = \alpha + \beta\, |\text{Sentiment}_t| + \gamma\, \text{Controls}_t + \varepsilon_t, \quad (4)$$

where Controls include the previous control variables and one-week-lagged stock market volatility.

Table 8 reports the results. Absolute GIFsentiment at day $t$ positively correlates with return volatility in the subsequent week ($t$ to $t$ + 5). In Column 1, a one-standard-deviation increase in absolute GIFsentiment corresponds to a 0.089 higher stock market volatility, significant at the 5% level. This represents 17% of the weekly volatility standard deviation of 0.524.[19] Results are similar in Column 2 when controlling for other sentiment proxies. The association between absolute GIFsentiment and subsequent one-month (*t* to *t* + 20) return volatility in Columns (3) and (4) is not statistically significant. This outcome is analogous to our return tests, which show a price decline following day *t* + 5; similarly, longer-horizon volatility declines. In untabulated predictive tests, we shift the volatility windows to begin on day *t* + 1. The results are quantitatively similar.

Absolute TEXTsentiment and absolute SELFDEC at day *t* are negatively correlated with stock market volatility over the subsequent week or month. These negative associations suggest these measures are correlated with fundamental news (see Section 2.4) that helps resolve uncertainty, reducing subsequent uncertainty.

Overall, our return volatility findings consistently suggest that visual-based GIFsentiment captures investor mood or attention-induced biases in expectations, leading to deviations of stock price from fundamentals and to excess volatility. In contrast, TEXTsentiment and SELFDEC appear to reflect fundamental information more than mood.

## 4 GIF Sentiment, GIF Disagreement, and Trading Activity

### 4.1 GIF sentiment and total trading volume

Behavioral models propose that sentiment shocks induce disagreement between rational and noise investors, increasing trading activity (De Long et al. 1990, Campbell, Grossman, and Wang 1993). If GIFsentiment reflects these dynamics, we expect extreme sentiment (high or low) to correlate with increased trading volume as the market absorbs these orders.

We test the association of GIFsentiment with market trading. Intraday level analysis is employed to enable more granular and timely examination of market reactions and their temporal alignment with sentiment shocks.

Table 9 examines the relation between aggregated 30-minute SPY trading volume and

[19] The magnitude is similar to that in Edmans et al. (2022). A one-standard-deviation increase in their music sentiment is associated with a 3.7 bps increase in stock market volatility, or 3.48% of the average weekly volatility of 1.06.

GIFsentiment. Following Tetlock (2007), we run the following regression:

$$\text{LogTotalVol}_{t+m\rightarrow t+n} = \alpha + \beta\, |\text{Sentiment}_t| + \delta\, \text{Controls}_t + \varepsilon_t, \quad (5)$$

where LogTotalVol represents the natural logarithm of one plus total SPY trading volume, cumulated within a given 30-minute interval $t$, and over subsequent one and two-day windows from $t$ + 30 minutes to $t$ + 1 day and from $t$ + 30 minutes to $t$ + 2 days respectively. $|\text{Sentiment}_t|$ is the standardized absolute value of GIFsentiment with mean of zero at 30-minute interval $t$, where high values indicate unusually strong positive or negative sentiment within that window.

Table 9 Panel A Columns 1 to 3 show positive coefficients for $|\text{Sentiment}_t|$ on the contemporaneous 30-minute interval $t$, over the next day ($t$ + 30m to $t$ + 1d), and over the second next day ($t$ + 30m to $t$ + 2d). This indicates both high and low GIFsentiment levels are associated with higher trading volume. A one-standard-deviation increase in GIFsentiment corresponds to 0.160 higher log trading volume contemporaneously, a 6.0% of its standard deviation (2.67), and 0.039 and 0.042 higher log volume over the next day and the second next day (2.7 and 2.8% of their standard deviation) respectively. Results are qualitatively similar when controlling for the five additional sentiment measures.

### 4.2 GIF disagreement and total trading volume

Belief heterogeneity models posit that disagreement among investors from differing expectations drives trading activity (Harrison and Kreps 1978; Kandel and Pearson 1995). Consistent with this, we test if GIFsentiment disagreement, measured as the dispersion or imbalance in bullish versus bearish declarations, correlates with increased trading volume. If GIFsentiment captures meaningful variation in investor beliefs, greater GIFsentiment disagreement should signal heightened divergence in expectations, resulting in higher trading volume.

We construct a measure of disagreement in GIF sentiment at each 30-minute interval as the standard deviation of raw GIF sentiment across user's posts within the 30-minute interval $t$. This measure is in the spirit of measures of analyst forecast dispersion and is similar to the standard deviation disagreement measure in Booker, Curtis, and Richardson (2023). We standardize this disagreement measure to have zero mean and unit variance.

We use this to estimate Equation (5) from Section 4.1 replacing $|\text{Sentiment}_t|$ with $\text{GIFDisagreement}_t$ and retaining the same controls. Table 9 Panel B presents these results. Columns 1 to 3 show positive $\text{GIFDisagreement}_\text{t}$ coefficients ($p < 0.01$), indicating that greater divergence in sentiment is associated with higher trading volume. A one-standard-deviation higher GIFDisagreement is associated with 0.116 higher log trading volume contemporaneously, which corresponds to 4.3% of the standard deviation of log volume (2.67). The same increase in GIFDisagreement is also associated with 0.070 and 0.068 higher log volume over the next day and the second day after, corresponding to 4.9% and 3.4% of the standard deviation of log volume at the corresponding horizons. This finding is consistent with GIFsentiment capturing meaningful heterogeneity in investor beliefs, with heightened disagreement corresponding to more active market trading.

## 5 GIF Sentiment and Fund Flows

If sentiment is associated with investor decisions, we expect to observe corresponding patterns in mutual fund investing. This consideration can be important because individual investors hold about 95% of long-term U.S. mutual fund total net assets (Investment Company Institute 2023). Since daily fund flows aggregate to the asset-class (Ivković and Weisbenner 2009), we investigate the predictive power of GIFsentiment for daily mutual fund flows between two asset classes, U.S. equities versus U.S. bonds. We test whether higher sentiment reflects greater investor optimism, which in turn encourages investors to shift toward equities relative to bonds.

Our daily equity and bond fund flow data are from EPFR Global, a private company that tracks performance and asset allocation of equity and debt mutual funds in developed and emerging markets.[20] Daily flows are computed as the ratio of dollar flow to fund total net assets (TNA). We estimate the following regression:

[20] As of 2024, EPFR global tracks over 151,000 equity share classes and 50,000 individual bonds, comprising more than $55 trillion in assets in developed, emerging and frontier markets (EPFR Product Overview 2024).

$$Y_{t+m \rightarrow t+n} = \alpha + \beta \, \text{Sentiment}_t + \gamma \, \text{Controls}_t + \varepsilon_t, \quad (6)$$

where $Y$ = EFF or BFF, which represent deseasonalized daily net equity and bond fund flows, respectively, scaled by TNA. Controls are as previously defined. To remove seasonality, equity and bond fund flows are first regressed on day-of-week and month-of-year indicators, with the residuals used as the dependent variables in Equation (6). Table 10 Panel A reports equity fund flow results and Panel B reports bond fund flow results.

Panel A shows that GIFsentiment is positively correlated with equity fund flows (EFF) in the subsequent week on days $t + 1$ to $t + 5$. A one-standard-deviation higher GIFsentiment is associated with 0.005 higher EFF during the subsequent week (20% of its EFF standard deviation). Results remain similar after including the five additional sentiment measures.

Panel B shows that GIFsentiment is negatively correlated with bond fund flows (BFF) on day 0 and during $t + 1$ to $t + 5$. With the five additional sentiment measures, GIFsentiment is unrelated to day 0 BFF but remains negatively correlated with subsequent week (days $t + 1$ to $t + 5$) BFF. A one-standard-deviation higher GIFsentiment corresponds to a 0.031 lower BFF during the subsequent week (17.4% of its standard deviation). In summary, high GIFsentiment forecasts inflows to equity funds and outflows from bond mutual funds in the subsequent week.

TEXTsentiment is negatively associated with equity fund flows and positively associated with bond fund flows. In contrast, SELFDEC and MEDIAsentiment are negatively associated with bond mutual fund flows, while MEDIAsentiment is also positively associated with equity fund flows on day $t$. Overall, both GIFsentiment and MEDIAsentiment exhibit a similar pattern, whereas high TEXTsentiment and SELFDEC do not show such behavior.

## 6 Conclusion

Investor sentiment is shaped by both individual reactions to traditional news media and social interactions among investors. So a comprehensive understanding of investor sentiment requires studying the communication dynamics between investors. A key window for such study is provided by data on behavior in social media platforms. Research in psychology and social media has emphasized the importance of multimodal communications, and especially to the expressive and attention-directing properties of dynamic visual content. This paper proposes a novel measure of investor sentiment derived from dynamic visuals, GIFs, shared by users on the social media platform Stocktwits.com.

Our daily measure of aggregate investor sentiment, GIFsentiment, is correlated with established exogenous mood proxies from past literature. Importantly, GIFsentiment is positively correlated with contemporaneous aggregate stock market returns and negatively predicts future market return at horizons of up to one month. GIFsentiment also predicts stock market volatility, trading activity, and shifts in mutual fund flows from bonds to equities. The return reversal pattern is consistent with transient sentiment-induced mispricing (De Long et al. 1990, Campbell, Grossman, and Wang 1993).

Our findings are robust to controlling for various sentiment proxies, including social media proxies such as self-declared sentiment and textual sentiment of postings; Baker-Wurgler sentiment; Michigan Consumer sentiment; and traditional mass media sentiment. They also hold when controlling for fundamentals such as U.S. economic policy uncertainty, U.S. macroeconomic activity, and past returns.

Consistent with investor sentiment and market mispricing predictions, the associations of GIF sentiment with returns are strongest for portfolios of small stocks and high idiosyncratic volatility stocks. These portfolios are usually viewed as more sensitive to retail investor misperceptions and costlier to arbitrage.

We find sharp differences between results for GIFsentiment and two alternative sentiment proxies, TEXTsentiment and SELFDEC sentiment. GIFsentiment outperforms both these sentiment measures in forecasting aggregate market return reversals. Unlike the return reversals predicted by GIFsentiment, neither TEXTsentiment nor SELFDEC is a significant predictor of stock returns in either the short or long run. This contrast is likely because dynamic visuals heavily capture emotion- or attention-driven biases in investor expectations, whereas text and user declarations reflect at least some meaningful fundamental information.

Our approach to measuring the sentiment of dynamic visual content analysis benefits from the large sample of Stocktwits.com participants who directly label the sentiment of their posts. Since participants on this platform have a strong interest in the stock market, our sentiment proxy is based on an ecologically relevant investment context as compared with the use other types of raters for estimating sentiment.

Our analysis of social media sentiment differs from most past studies aggregate market return predictability in predicting reversals at high frequencies. Most past studies identify reversals in aggregate market returns at long time horizons of up to several years (Fama and French 1988, Poterba and Summers 1988). In contrast, GIF sentiment predicts negative aggregate market returns at daily to weekly frequencies.

In conclusion, this paper is the first to study GIFs, a type of dynamic visual representations, as a means of communicating ideas about stock investing with other investors. It exploits dynamic visuals to construct a high-frequency (daily) market sentiment measure that predicts aggregate stock market returns and other aggregate market outcomes. As such, the paper contributes new insights about investor sentiment, stock market return predictability, social media in financial markets, and the growing field of social economics and finance.

## Figure 1
## Examples of Key Static Frames from Animated GIFs

This figure presents the key static frames from four example GIFs. The corresponding animated GIFs can be viewed via the following hyperlinks: (a) stylized alien figure with glitch effects (GiphyID: l3q2Lu62MjV4N68bm); (b) illustrated bear with declining trend line indicating feeling bearish (GiphyID: UfX4XeBMXWmNoGvBVK); (c) Tom Hanks expressing squeezing using orange (GiphyID: 3o7TKPdUkkbCAVqWk0); and (d) two bears waving their claws (GiphyID: 13XarhksGkhCZG).

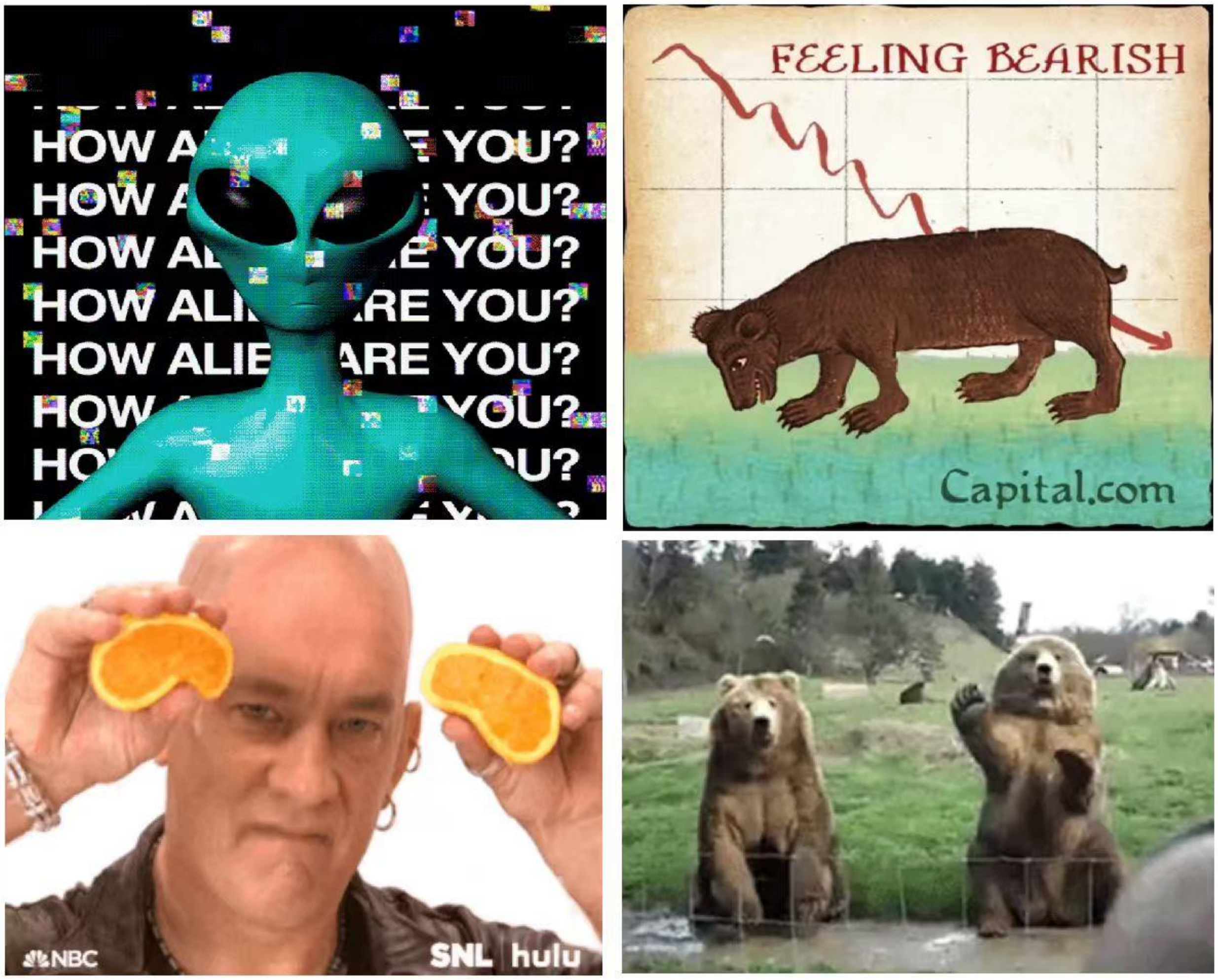

**Figure 2**
**The Trend of GIF Postings on Stocktwits Over Time**

This figure plots the time trend of GIF postings on Stocktwits from September 1, 2020, to October 31, 2024. Beginning on September 1, 2020, Stocktwits started supporting GIFs posted from all users' accounts.

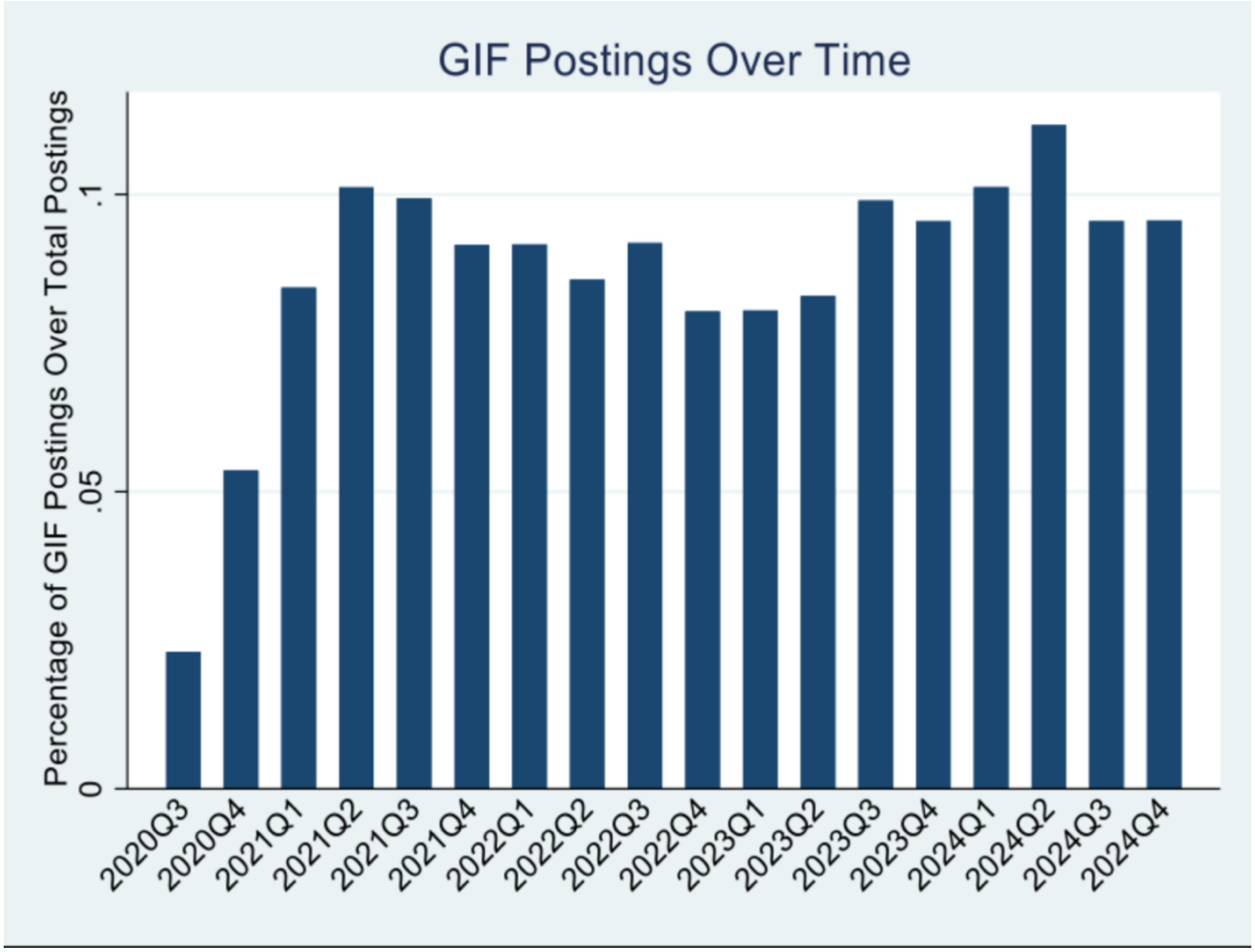

**Figure 3**
**The Time Series of Alternative Sentiment Measures**

Panel A shows the time series of daily GIF sentiment, SELFDEC, Text sentiment, and Media sentiment for our sample period between September 1, 2020, and October 31, 2024. Panel B shows the time series of monthly GIF sentiment, BW sentiment, and ICS for our sample period.

Panel A Time-Series of Daily Sentiment Measures

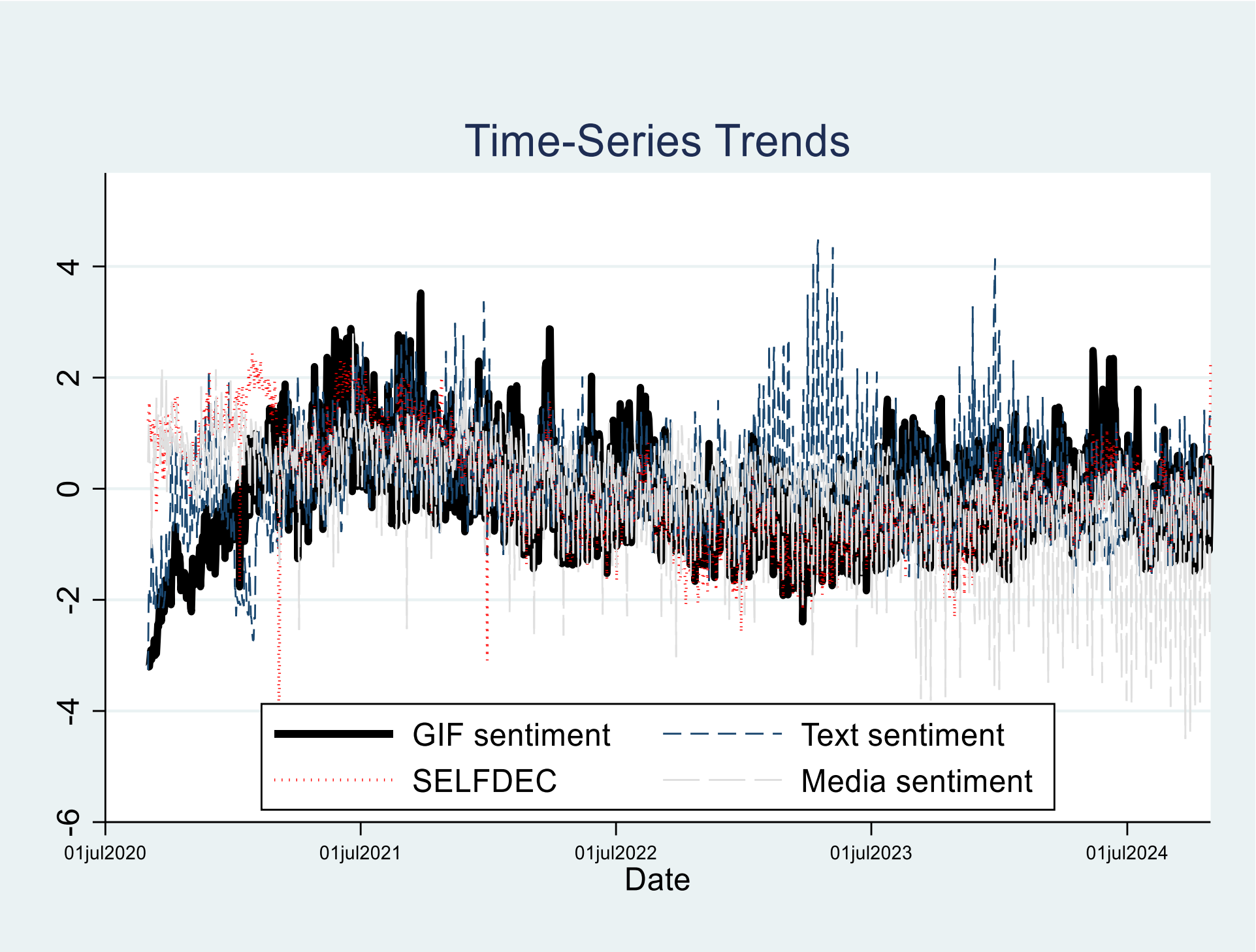

Panel B Time-Series of Monthly Sentiment Measures

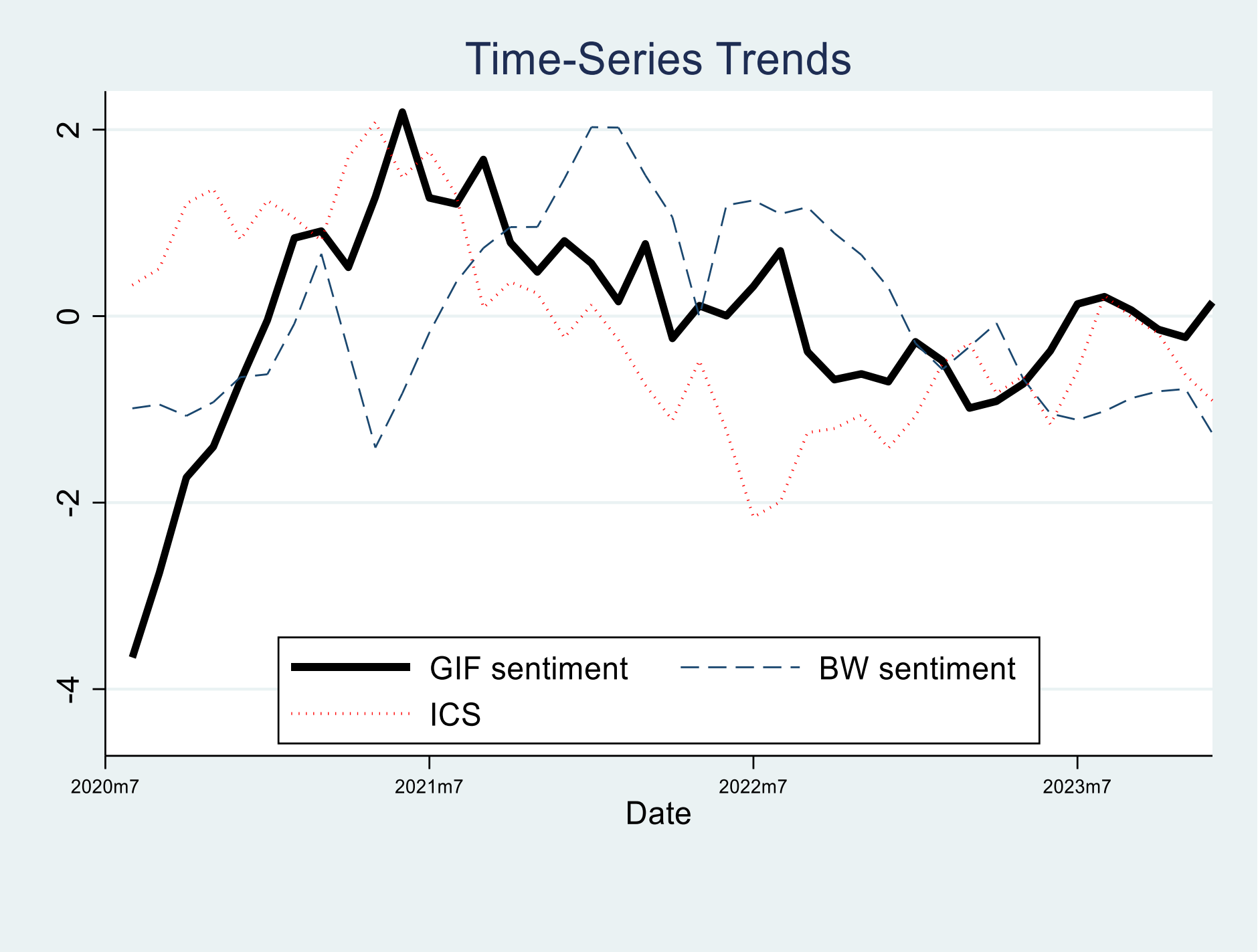

**Table 1**
**GIF and Bullish/Bearish Self-Declarations**

Panel A presents the distribution of the number of bullish declarations (#Declared_Bullish), bearish declarations (#Declared_Bearish), and the frequency of each unique GIF (#Appearance) at the GIF-day level. These statistics are calculated using only information available at the daily level to avoid look-ahead bias. Panel B reports the autocorrelation statistics of GIFsentiment, conditional on different minimum appearance thresholds, to assess its temporal persistence across GIFs with varying levels of appearances.

Panel A: Distribution of Bullish and Bearish Declarations at Daily-GIF Level for All GIFs

| Variable (N= 4,121,312) | Mean | Std Dev | p10 | p25 | p50 | p75 | p90 |
|---|---|---|---|---|---|---|---|
| #Declared_Bullish | 127.57 | 452.96 | 1 | 3 | 15 | 74 | 268 |
| #Declared_Bearish | 16.15 | 76.81 | 0 | 0 | 1 | 6 | 28 |
| #Appearance | 184.35 | 582.64 | 1 | 5 | 25 | 123 | 424 |

Panel B: Autocorrelation of GIF Sentiment

| #Appearance | N | Mean | Std Dev | P10 | P25 | P50 | P75 | P90 |
|---|---|---|---|---|---|---|---|---|
| ≥ 5 | 94,884 | 0.59 | 0.39 | 0.01 | 0.43 | 0.72 | 0.88 | 0.95 |
| ≥ 10 | 59,799 | 0.69 | 0.30 | 0.25 | 0.57 | 0.79 | 0.91 | 0.96 |
| ≥ 25 | 23,834 | 0.79 | 0.21 | 0.51 | 0.72 | 0.86 | 0.94 | 0.97 |

**Table 2**
**Summary Statistics**

Panel A reports the summary statistics for GIFsentiment, TEXTsentiment, self-declared sentiment (SELFDEC), Baker-Wurgler sentiment index (BW), consumer sentiment index (ICS), news media sentiment (MEDIAsentiment), and the daily returns on the CRSP S&P 500 Index (SPX). Panel B reports the pairwise Pearson correlations between the sentiment measures. Values in bold indicate *p*-values < 0.01, while non-bold values correspond to *p*-values > 0.10. GIFsentiment is calculated following Equation (1) in the text. All variable definitions are in Table A2.

Panel A: Summary Statistics of Main Variables

| Variable | N | Mean | Std Dev | P10 | P25 | P50 | P75 | P90 |
|---|---|---|---|---|---|---|---|---|
| GIFsentiment | 1,045 | 0.05 | 0.01 | 0.03 | 0.03 | 0.04 | 0.05 | 0.06 |
| TEXTsentiment | 1,045 | 0.09 | 0.02 | 0.06 | 0.07 | 0.08 | 0.09 | 0.10 |
| SELFDEC | 1,045 | 0.28 | 0.08 | 0.15 | 0.17 | 0.22 | 0.27 | 0.34 |
| BW | 858 | 0.78 | 0.78 | -0.24 | -0.12 | -0.03 | 0.89 | 1.25 |
| ICS (consumer index) | 1,045 | 69.44 | 9.03 | 56.70 | 58.40 | 62.80 | 68.20 | 76.90 |
| MEDIAsentiment | 1,045 | 0.11 | 0.04 | 0.05 | 0.06 | 0.08 | 0.11 | 0.14 |
| SPX | 1,045 | 0.05 | 1.06 | -1.70 | -1.20 | -0.51 | 0.07 | 0.70 |

Panel B: Pearson Correlations Between Sentiment Variables

| | GIFsentiment | TEXTsentiment | SELFDEC | BW | ICS | MEDIAsentiment |
|---|---|---|---|---|---|---|
| GIFsentiment | 1 | | | | | |
| TEXTsentiment | **0.45** | 1 | | | | |
| SELFDEC | **0.49** | **0.11** | 1 | | | |
| BW | **0.47** | **0.34** | **0.26** | 1 | | |
| ICS | **0.16** | -0.04 | **0.60** | 0.02 | 1 | |
| MEDIAsentiment | **0.16** | **0.17** | **0.53** | **0.34** | **0.33** | 1 |

**Table 3**
**Correlation Between Sentiment Measures and Mood Proxies**

Panel A reports the Pearson correlation of sentiment measures and **daily** mood proxies. DCC is the average daily cloud cover, deseasonalized by each week's average cloud cover. ΔCOVID Index is the change in daily containment and closure index. Panel B reports the Pearson correlation of sentiment measures and **monthly** mood proxies. Positive months is an indicator variable that equals 1 in January and March and 0 otherwise. Negative months is an indicator variable that equals 1 in September and October and 0 otherwise. Panel C reports the Pearson correlation of sentiment measures and an information proxy, %PositiveEANews, measured by the percentage of earnings news that meet or beat analyst consensus forecast on the announcement date. GIFsentiment is the daily appearance-weighted average sentiment of GIFs posted on Stocktwits. BW and ICS sentiment measures are monthly measures so all correlations with BW and ICS are estimated monthly. The other variables in other columns are daily. All sentiment measures are standardized to have zero mean and unit variance. *, **, and *** denote significance at the 10%, 5%, and 1% level, respectively. Variable definitions are in Table A2.

Panel A: The Pearson Correlation Between Sentiment Measures and Daily Mood Proxies

| | (1) | (2) | (3) | (4) | (5) | (6) |
|---|---|---|---|---|---|---|
| VARIABLES | GIFsentiment | TEXTsentiment | SELFDEC | BW | ICS | MEDIAsentiment |
| DCC | -0.118*** | -0.047 | -0.030 | -0.006 | 0.062 | 0.114*** |
| | (0.000) | (0.129) | (0.331) | (0.968) | (0.671) | (0.000) |
| ΔCOVID Index | -0.133*** | -0.057 | -0.092** | -0.233 | -0.480*** | -0.008 |
| | (0.001) | (0.165) | (0.027) | (0.232) | (0.010) | (0.846) |

Panel B: The Pearson Correlation Between Sentiment Measures and Monthly Mood Proxies

| | (1) | (2) | (3) | (4) | (5) | (6) |
|---|---|---|---|---|---|---|
| VARIABLES | GIFsentiment | TEXTsentiment | SELFDEC | BW | ICS | MEDIAsentiment |
| Positive Months | 0.037 | -0.060** | 0.012 | 0.040 | 0.053 | 0.011 |
| | (0.235) | (0.055) | (0.698) | (0.802) | (0.715) | (0.721) |
| Negative Months | -0.222*** | -0.182*** | 0.003 | -0.069 | -0.026 | 0.037 |
| | (0.000) | (0.000) | (0.922) | (0.666) | (0.858) | (0.229) |

Panel C: The Pearson Correlation Between Information Proxy and Six Sentiment Measures

| | (1) | (2) | (3) | (4) | (5) | (6) |
|---|---|---|---|---|---|---|
| VARIABLES | GIFsentiment | TEXTsentiment | SELFDEC | BW | ICS | MEDIAsentiment |
| %PositiveEANews | -0.015 | -0.066** | 0.104*** | -0.056 | 0.343** | 0.254*** |
| | (0.621) | (0.032) | (0.000) | (0.725) | (0.014) | (0.000) |

**Table 4**
**Regressions of S&P 500 Index Returns on the Sentiment Indices**

This table reports the regression estimates of Equation (3) from September 2020 to October 2024. The dependent variable is the Standard and Poor's 500 Index (SPX) return at alternative windows. We multiply returns by 100 to interpret coefficients as percentage points. The main independent variable, GIFsentiment is the daily appearance-weighted average sentiment of GIFs posted on Stocktwits. Sentiment measures are standardized to have zero mean and unit variance. Standard errors (reported in parentheses) are computed using a moving block bootstrap as described in the main text. *, **, and *** denote significance at the 10%, 5%, and 1% level, respectively. Variable definitions are in Table A2.

Panel A: GIF Sentiment Alone

| VARIABLES | (1)<br>Ret(*t*) | (2)<br>Ret[*t* + 1, *t* + 5] | (3)<br>Ret[*t* + 1, *t* + 20] |
|---|---|---|---|
| GIFsentiment | 0.273*** | 0.035 | -1.265*** |
| | (0.066) | (0.155) | (0.352) |
| EPU | 0.001 | 0.005** | 0.003 |
| | (0.001) | (0.002) | (0.003) |
| ADS | -0.031 | 0.776** | 2.076*** |
| | (0.112) | (0.324) | (0.803) |
| Ret(*t*) | | -0.139 | 0.007 |
| | | (0.115) | (0.248) |
| Ret[*t* - 5, *t* - 1] | -0.054** | -0.107* | -0.030 |
| | (0.026) | (0.056) | (0.171) |
| Ret[*t* - 21, *t* - 6] | 0.006 | -0.048 | -0.249** |
| | (0.016) | (0.041) | (0.121) |
| Log#EA | -0.004 | 0.118 | -0.366 |
| | (0.045) | (0.110) | (0.273) |
| Log#AbnMessages | -0.023 | -0.082 | 0.952 |
| | (0.135) | (0.431) | (0.744) |
| Observations | 1,007 | 1,002 | 988 |
| Adjusted R-squared | 0.039 | 0.049 | 0.139 |

Panel B: Six Sentiment Measures

| VARIABLES | (1)<br>Ret(*t*) | (2)<br>Ret[*t* + 1, *t* + 5] | (3)<br>Ret[*t* + 1, *t* + 20] |
|---|---|---|---|
| GIFsentiment | 0.291** | -0.007 | -1.172** |
| | (0.120) | (0.282) | (0.584) |
| TEXTsentiment | 0.150* | -0.047 | 0.638 |
| | (0.077) | (0.205) | (0.390) |
| SELFDEC | 0.003 | -0.519 | -0.148 |
| | (0.139) | (0.319) | (0.705) |
| BW | -0.334*** | -0.435* | -1.665** |
| | (0.096) | (0.222) | (0.749) |
| ICS | -0.287*** | 0.357 | 0.847 |
| | (0.095) | (0.264) | (0.892) |
| MEDIAsentiment | 0.537*** | 0.146 | -0.022 |
| | (0.120) | (0.178) | (0.387) |
| EPU | 0.001* | 0.003 | 0.002 |
| | (0.001) | (0.002) | (0.003) |
| ADS | 0.032 | 1.001** | 1.941** |
| | (0.137) | (0.401) | (0.795) |
| Ret(t) | | -0.087 | -0.309 |
| | | (0.155) | (0.253) |
| Ret[*t* - 5, *t* - 1] | -0.058** | -0.190** | -0.310* |
| | (0.029) | (0.079) | (0.164) |
| Ret[*t* - 21, *t* - 6] | -0.016 | -0.125* | -0.249* |
| | (0.020) | (0.068) | (0.136) |
| Log#EA | 0.072 | -0.018 | 0.341 |
| | (0.054) | (0.125) | (0.248) |
| Log#AbnMessages | -0.124 | -0.273 | 0.039 |
| | (0.190) | (0.475) | (0.922) |
| Observations | 822 | 822 | 822 |
| Adjusted R-squared | 0.209 | 0.140 | 0.364 |

**Table 5**
**Regressions of S&P 500 Index Returns on Positive and Negative GIF Sentiment**

This table reports regression estimates from Equation (3), estimated over the period from September 2020 to October 2024, where GIFsentiment is replaced with its two components: PositiveGIFsentiment and NegativeGIFsentiment. PositiveGIFsentiment is constructed using only GIFs with a net positive sentiment score, while NegativeGIFsentiment is based on GIFs with a net negative sentiment score. The dependent variable is the Standard and Poor's 500 Index (SPX) return at alternative windows. We multiply returns by 100 to interpret coefficients as percentage points. Sentiment measures are standardized to have zero mean and unit variance. Standard errors (reported in parentheses) are computed using a moving block bootstrap as described in the main text. *, **, and *** denote significance at the 10%, 5%, and 1% level, respectively. Variable definitions are in Table A2. Table 5 has one fewer observation than Table 4 because, on one trading day with GIFsentiment equal to zero, both PositiveGIFsentiment and NegativeGIFsentiment are undefined and the observation is dropped.

Panel A: GIF Sentiment Alone

| VARIABLES | (1)<br>Ret(*t*) | (2)<br>Ret[*t* + 1, *t* + 5] | (3)<br>Ret[*t* + 1, *t* + 20] |
|---|---|---|---|
| PositiveGIFsentiment | 0.242*** | 0.086 | -0.878** |
| | (0.072) | (0.193) | (0.405) |
| NegativeGIFsentiment | 0.155*** | -0.409** | -0.489 |
| | (0.053) | (0.170) | (0.457) |
| EPU | 0.001 | 0.002 | 0.005* |
| | (0.001) | (0.002) | (0.003) |
| ADS | -0.086 | 0.788** | 2.106** |
| | (0.127) | (0.368) | (0.850) |
| Ret(*t*) | | -0.072 | -0.001 |
| | | (0.139) | (0.270) |
| Ret[*t* - 5, *t* - 1] | -0.059** | -0.057 | -0.165 |
| | (0.025) | (0.062) | (0.135) |
| Ret[*t* - 21, *t* - 6] | 0.008 | -0.003 | -0.125 |
| | (0.017) | (0.048) | (0.126) |
| Log#EA | -0.010 | 0.328*** | 0.310 |
| | (0.046) | (0.116) | (0.231) |
| Log#AbnMessages | 0.074 | -0.186 | 0.269 |
| | (0.161) | (0.401) | (0.789) |
| Observations | 1,006 | 1,001 | 987 |
| Adjusted R-squared | 0.070 | 0.067 | 0.154 |

Panel B: Six Sentiment Measures

| VARIABLES | (1)<br>Ret(*t*) | (2)<br>Ret[*t* + 1, *t* + 5] | (3)<br>Ret[*t* + 1, *t* + 20] |
|---|---|---|---|
| PositiveGIFsentiment | 0.435*** | 0.569** | -1.588*** |
| | (0.130) | (0.252) | (0.607) |
| NegativeGIFsentiment | 0.514*** | -0.067 | 0.147 |
| | (0.166) | (0.363) | (0.841) |
| TEXTsentiment | 0.065 | -0.301 | 0.126 |
| | (0.082) | (0.216) | (0.419) |
| SELFDEC | -0.241 | -0.076 | 1.369* |
| | (0.185) | (0.331) | (0.797) |
| BW | -0.598*** | -0.625** | -1.099* |
| | (0.107) | (0.266) | (0.563) |
| ICS | -0.257*** | 0.004 | 0.208 |
| | (0.090) | (0.277) | (0.774) |
| MEDIAsentiment | 0.516*** | -0.090 | -0.729* |
| | (0.126) | (0.217) | (0.401) |
| EPU | 0.000 | 0.006** | 0.005* |
| | (0.001) | (0.003) | (0.003) |
| ADS | 0.081 | 1.043*** | 2.149*** |
| | (0.134) | (0.392) | (0.757) |
| Ret(*t*) | | -0.200 | -0.255 |
| | | (0.152) | (0.254) |
| Ret[*t* - 5, *t* - 1] | -0.069** | -0.167** | -0.324** |
| | (0.029) | (0.065) | (0.142) |
| Ret[*t* - 21, *t* - 6] | -0.050** | -0.082* | -0.366*** |
| | (0.019) | (0.047) | (0.134) |
| Log#EA | 0.032 | 0.166 | -0.076 |
| | (0.051) | (0.126) | (0.234) |
| Log#AbnMessages | -0.082 | -0.268 | 0.405 |
| | (0.196) | (0.607) | (0.851) |
| Observations | 821 | 821 | 821 |
| Adjusted R-squared | 0.248 | 0.098 | 0.258 |

**Table 6**
**Intraday GIF Sentiment and Stock Returns**

This table reports the regression estimates of Equation (3) from September 2020 to December 2023, using intraday return measures. The dependent variable is the Standard and Poor's ETF (SPY) return at alternative windows. We multiply returns by 100 to interpret coefficients as percentage points. The main independent variable, GIFsentiment is measured at the 30-minute interval level per trading day. Sentiment measures are standardized to have zero mean and unit variance. Standard errors (reported in parentheses) are computed using a moving block bootstrap as described in the main text. *, **, and *** denote significance at the 10%, 5%, and 1% level, respectively. Variable definitions are in Table A2.

Panel A: GIF Sentiment Alone

| VARIABLES | (1)<br>Ret(*t*) | (2)<br>Ret[*t* + 30m, *t* + 1d] | (3)<br>Ret[*t* + 30m, *t* + 2d] |
|---|---|---|---|
| GIFsentiment | 0.057*** | -0.031 | -0.077** |
| | (0.005) | (0.024) | (0.037) |
| EPU | 0.000 | -0.001 | -0.000 |
| | (0.000) | (0.000) | (0.001) |
| ADS | -0.010 | 0.134*** | 0.343*** |
| | (0.007) | (0.046) | (0.080) |
| Ret(*t*) | | 0.088 | 0.100 |
| | | (0.082) | (0.120) |
| Ret[*t* - 5, *t* - 1] | -0.005 | -0.006 | 0.042 |
| | (0.010) | (0.047) | (0.068) |
| Ret[*t* - 21, *t* - 6] | -0.009* | -0.021 | -0.070 |
| | (0.005) | (0.033) | (0.052) |
| Log#EA | 0.002 | 0.000 | 0.000 |
| | (0.003) | (0.013) | (0.020) |
| Log#AbnMessages | -0.039*** | -0.006 | -0.005 |
| | (0.011) | (0.067) | (0.109) |
| Observations | 10,493 | 10,481 | 10,468 |
| Adjusted R-squared | 0.043 | 0.009 | 0.025 |

Panel B: Six Sentiment Measures

| VARIABLES | (1)<br>Ret(*t*) | (2)<br>Ret[*t* + 30m, *t* + 1d] | (3)<br>Ret[*t* + 30m, *t* + 2d] |
|---|---|---|---|
| GIFsentiment | 0.016*** | -0.017 | -0.073* |
| | (0.006) | (0.029) | (0.043) |
| TEXTsentiment | 0.017*** | 0.009 | 0.059 |
| | (0.006) | (0.027) | (0.045) |
| SELFDEC | 0.081*** | -0.021 | 0.085 |
| | (0.009) | (0.044) | (0.076) |
| BW | -0.027*** | -0.008 | -0.037 |
| | (0.006) | (0.038) | (0.072) |
| ICS | -0.031*** | 0.000 | -0.059 |
| | (0.006) | (0.032) | (0.062) |
| MEDIAsentiment | -0.007 | -0.014 | -0.068* |
| | (0.004) | (0.026) | (0.041) |
| EPU | -0.000 | -0.001 | -0.001 |
| | (0.000) | (0.000) | (0.001) |
| ADS | -0.011 | 0.144*** | 0.320*** |
| | (0.008) | (0.050) | (0.099) |
| Ret(t) | | 0.089 | -0.110 |
| | | (0.089) | (0.130) |
| Ret[*t* - 5, *t* - 1] | -0.023** | -0.006 | -0.084 |
| | (0.011) | (0.050) | (0.076) |
| Ret[*t* - 21, *t* - 6] | -0.015*** | -0.021 | -0.013 |
| | (0.006) | (0.032) | (0.055) |
| Log#EA | 0.004 | 0.000 | -0.005 |
| | (0.003) | (0.013) | (0.021) |
| Log#AbnMessages | -0.058*** | 0.013 | 0.103 |
| | (0.016) | (0.075) | (0.141) |
| Observations | 10,493 | 10,481 | 10,468 |
| Adjusted R-squared | 0.083 | 0.010 | 0.027 |

**Table 7**
**GIF Sentiment and Limits to Arbitrage**

This table reports the regression estimates of Equation (3) from September 2020 to October 2024. Panel A dependent variables are the value-weighted daily returns for the small (bottom quintile) and large (top quintile) cap portfolios, sorted based on market capitalization of the firm. Panel B dependent variables are the value-weighted daily returns for the top and bottom quintile portfolios, sorted based on idiosyncratic volatility using the Fama and French (1993) three factors and Cahart (1997) momentum factor. Panel C dependent variables are the value-weighted daily returns for the top and bottom quintile portfolios, sorted based on total return volatility. We multiply the returns by 100 so coefficients are interpreted as percentage points. The main independent variable, GIFsentiment is the daily appearance-weighted average sentiment of GIFs posted on Stocktwits. Sentiment measures are standardized to have zero mean and unit variance. Standard errors (reported in parentheses) are computed using a moving block bootstrap as described in the main text. *, **, and *** denote significance at the 10%, 5%, and 1% level, respectively. Variable definitions are in Table A2.

Panel A1: Small vs. Large Cap Index Returns; GIF Sentiment Alone

| | Small Cap | | | Large Cap | | |
|---|---|---|---|---|---|---|
| | (1) | (2) | (3) | (4) | (5) | (6) |
| VARIABLES | Ret(*t*) | Ret[*t* + 1, *t* + 5] | Ret[*t* + 1, *t* + 20] | Ret(*t*) | Ret[*t* + 1, *t* + 5] | Ret[*t* + 1, *t* + 20] |
| GIFsentiment | 0.657*** | -0.213 | -3.043* | 0.287*** | -0.219 | -1.372*** |
| | (0.140) | (0.427) | (1.623) | (0.066) | (0.163) | (0.369) |
| Controls | YES | YES | YES | YES | YES | YES |
| Observations | 1,007 | 1,002 | 988 | 1,007 | 1,002 | 988 |
| Adjusted R-squared | 0.049 | 0.034 | 0.110 | 0.051 | 0.038 | 0.144 |
| (Small-Large) t-stats | 2.391 | 0.013 | -1.004 | | | |

Panel A2: Small vs. Large Cap Index Returns; Six Sentiment Measures

| | Small Cap | | | Large Cap | | |
|---|---|---|---|---|---|---|
| | (1) | (2) | (3) | (4) | (5) | (6) |
| VARIABLES | Ret(*t*) | Ret[*t* + 1, *t* + 5] | Ret[*t* + 1, *t* + 20] | Ret(*t*) | Ret[*t* + 1, *t* + 5] | Ret[*t* + 1, *t* + 20] |
| GIFsentiment | 0.381** | -0.772 | -4.522** | 0.193* | -0.033 | -1.766*** |
| | (0.190) | (0.822) | (1.928) | (0.111) | (0.278) | (0.676) |
| TEXTsentiment | 0.038 | -0.794 | -1.865 | 0.220*** | -0.214 | 0.216 |
| | (0.135) | (0.701) | (1.613) | (0.085) | (0.194) | (0.445) |
| SELFDEC | 0.495* | 1.942 | 4.937** | 0.293* | 0.169 | 1.680** |
| | (0.275) | (1.258) | (2.489) | (0.150) | (0.323) | (0.769) |
| BW | -0.599*** | -1.080* | -3.083** | -0.472*** | -0.391* | -1.335** |
| | (0.149) | (0.580) | (1.555) | (0.109) | (0.215) | (0.594) |
| ICS | -0.517** | -0.670 | -1.129 | -0.366*** | -0.364 | 0.041 |
| | (0.218) | (0.745) | (1.718) | (0.115) | (0.333) | (0.793) |
| MEDIAsentiment | 0.361** | -0.085 | 0.876 | 0.418*** | 0.032 | -0.552 |
| | (0.150) | (0.398) | (0.942) | (0.120) | (0.216) | (0.408) |
| Controls | YES | YES | YES | YES | YES | YES |
| Observations | 822 | 822 | 822 | 822 | 822 | 822 |
| Adjusted R-squared | 0.166 | 0.150 | 0.272 | 0.209 | 0.065 | 0.276 |
| (Small-Large) t-stats | 0.854 | -0.852 | -1.349 | | | |

Panel B1: High vs. Low Idiosyncratic Volatility of Returns; GIF Sentiment Alone

| | High Idiosyncratic Volatility | | | Low Idiosyncratic Volatility | | |
|---|---|---|---|---|---|---|
| | (1) | (2) | (3) | (4) | (5) | (6) |
| VARIABLES | Ret($t$) | Ret[$t + 1$, $t + 5$] | Ret[$t + 1$, $t + 20$] | Ret($t$) | Ret[$t + 1$, $t + 5$] | Ret[$t + 1$, $t + 20$] |
| GIFsentiment | 0.715*** | -0.584 | -3.800*** | 0.130*** | -0.190* | -0.962*** |
| | (0.131) | (0.415) | (1.194) | (0.034) | (0.097) | (0.212) |
| Controls | YES | YES | YES | YES | YES | YES |
| Observations | 1,007 | 1,002 | 988 | 1,007 | 1,002 | 988 |
| Adjusted R-squared | 0.097 | 0.037 | 0.138 | 0.044 | 0.049 | 0.186 |
| (High-Low) t-stats | 4.322 | -0.924 | -2.340 | | | |

Panel B2: High vs. Low Idiosyncratic Volatility of Returns; Six Sentiment Measures

| | High Idiosyncratic Volatility | | | Low Idiosyncratic Volatility | | |
|---|---|---|---|---|---|---|
| | (1) | (2) | (3) | (4) | (5) | (6) |
| VARIABLES | Ret($t$) | Ret[$t + 1$, $t + 5$] | Ret[$t + 1$, $t + 20$] | Ret($t$) | Ret[$t + 1$, $t + 5$] | Ret[$t + 1$, $t + 20$] |
| GIFsentiment | 0.445* | -0.471 | -5.759*** | 0.128* | -0.111 | -1.228*** |
| | (0.241) | (0.810) | (2.149) | (0.074) | (0.182) | (0.347) |
| TEXTsentiment | 0.171 | -1.208*** | -2.879* | 0.072* | -0.199** | -0.204 |
| | (0.171) | (0.457) | (1.576) | (0.041) | (0.097) | (0.239) |
| SELFDEC | 0.797*** | 0.163 | 4.371* | 0.019 | 0.141 | 0.943** |
| | (0.276) | (1.141) | (2.454) | (0.091) | (0.189) | (0.407) |
| BW | -0.892*** | 0.070 | -1.670 | -0.195*** | -0.072 | -0.690** |
| | (0.155) | (0.617) | (1.886) | (0.049) | (0.136) | (0.322) |
| ICS | -0.805*** | 0.084 | -0.419 | -0.143*** | 0.033 | 0.110 |
| | (0.156) | (0.743) | (2.165) | (0.046) | (0.151) | (0.384) |
| MEDIAsentiment | 0.675*** | 0.097 | -0.866 | 0.267*** | -0.074 | -0.067 |
| | (0.178) | (0.429) | (1.085) | (0.058) | (0.109) | (0.246) |
| Controls | YES | YES | YES | YES | YES | YES |
| Observations | 822 | 822 | 822 | 822 | 822 | 822 |
| Adjusted R-squared | 0.234 | 0.059 | 0.344 | 0.213 | 0.037 | 0.283 |
| (High-Low) t-stats | 1.257 | -0.434 | -2.081 | | | |

Panel C1: High vs. Low Total Return Volatility; GIF Sentiment Alone

| | High Total Return Volatility | | | Low Total Return Volatility | | |
|---|---|---|---|---|---|---|
| | (1) | (2) | (3) | (4) | (5) | (6) |
| VARIABLES | Ret(*t*) | Ret[*t* + 1, *t* + 5] | Ret[*t* + 1, *t* + 20] | Ret(*t*) | Ret[*t* + 1, *t* + 5] | Ret[*t* + 1, *t* + 20] |
| GIFsentiment | 0.739*** | -0.235 | -3.060** | 0.052*** | -0.025 | -0.717*** |
| | (0.148) | (0.486) | (1.316) | (0.018) | (0.080) | (0.188) |
| Controls | YES | YES | YES | YES | YES | YES |
| Observations | 1,007 | 1,002 | 988 | 1,007 | 1,002 | 988 |
| Adjusted R-squared | 0.093 | 0.011 | 0.121 | 0.046 | 0.052 | 0.177 |
| (High-Low) t-stats | 4.608 | -0.426 | -1.763 | | | |

Panel C2: High vs. Low Total Return Volatility; Six Sentiment Measures

| | High Total Return Volatility | | | Low Total Return Volatility | | |
|---|---|---|---|---|---|---|
| | (1) | (2) | (3) | (4) | (5) | (6) |
| VARIABLES | Ret(*t*) | Ret[*t* + 1, *t* + 5] | Ret[*t* + 1, *t* + 20] | Ret(*t*) | Ret[*t* + 1, *t* + 5] | Ret[*t* + 1, *t* + 20] |
| GIFsentiment | 0.481** | -0.287 | -5.619*** | 0.049 | 0.230 | -0.622* |
| | (0.218) | (0.764) | (2.008) | (0.030) | (0.145) | (0.363) |
| TEXTsentiment | 0.210 | -1.269*** | -3.778*** | 0.024 | -0.143** | -0.153 |
| | (0.170) | (0.445) | (1.449) | (0.022) | (0.068) | (0.197) |
| SELFDEC | 0.806*** | 0.901 | 5.474** | 0.021 | -0.228 | 0.058 |
| | (0.303) | (0.889) | (2.564) | (0.036) | (0.146) | (0.419) |
| BW | -1.025*** | -0.644 | -0.237 | -0.061** | -0.054 | -0.421 |
| | (0.157) | (0.621) | (1.869) | (0.025) | (0.077) | (0.278) |
| ICS | -0.711*** | -0.315 | 0.038 | -0.066*** | 0.087 | 0.879*** |
| | (0.185) | (0.677) | (2.156) | (0.022) | (0.088) | (0.313) |
| MEDIAsentiment | 0.632*** | 0.117 | -0.496 | 0.062** | 0.030 | -0.046 |
| | (0.147) | (0.431) | (1.123) | (0.025) | (0.079) | (0.170) |
| Controls | YES | YES | YES | YES | YES | YES |
| Observations | 822 | 822 | 822 | 822 | 822 | 822 |
| Adjusted R-squared | 0.260 | 0.134 | 0.376 | 0.141 | 0.105 | 0.427 |
| (High-Low) t-stats | 1.963 | -0.665 | -2.449 | | | |

**Table 8**
**Regressions of Stock Market Volatility on the Sentiment Proxies**

This table reports the regression estimates of Equation (4) from September 2020 to October 2024. The dependent variable, Volatility% is the standard deviation of daily S&P 500 Index return over two windows: from day *t* through *t* + 5 and from day *t* through *t* + 20. The main independent variable, |GIFsentiment| is the absolute value of daily appearance-weighted average sentiment of GIFs posted on Stocktwits. Standard errors (reported in parentheses) are computed using a moving block bootstrap as described in the main text. *, **, and *** denote significance at the 10%, 5%, and 1% level, respectively. Variable definitions are in Table A2.

| VARIABLES | (1) Volatility[*t*, *t* + 5](%) | (2) Volatility [*t*, *t* + 5](%) | (3) Volatility [*t*, *t* + 20](%) | (4) Volatility [*t*, t + 20](%) |
|---|---|---|---|---|
| \|GIFsentiment\| | 0.089** | 0.123** | 0.037 | 0.044 |
| | (0.039) | (0.050) | (0.033) | (0.039) |
| \|TEXTsentiment\| | | -0.080** | | -0.064*** |
| | | (0.031) | | (0.023) |
| \|SELFDEC\| | | -0.123* | | -0.043 |
| | | (0.066) | | (0.047) |
| \|BW\| | | -0.074 | | -0.033 |
| | | (0.068) | | (0.063) |
| \|ICS\| | | -0.014*** | | -0.010*** |
| | | (0.003) | | (0.002) |
| \|MEDIAsentiment\| | | 0.016 | | -0.021 |
| | | (0.043) | | (0.039) |
| EPU | 0.000 | -0.000 | 0.000 | -0.000 |
| | (0.000) | (0.000) | (0.000) | (0.000) |
| ADS | -0.014 | -0.065 | -0.025 | -0.061 |
| | (0.044) | (0.047) | (0.042) | (0.046) |
| Ret[*t* - 5, *t* - 1] | -0.061*** | -0.065*** | -0.028*** | -0.031*** |
| | (0.013) | (0.014) | (0.010) | (0.011) |
| Ret[*t* - 21, *t* - 6] | -0.019** | -0.024*** | -0.014** | -0.015* |
| | (0.008) | (0.009) | (0.007) | (0.008) |
| Volatility%[*t* - 5, *t* - 1] | 0.251*** | 0.237*** | 0.281*** | 0.292*** |
| | (0.060) | (0.070) | (0.044) | (0.054) |
| Log#EA | 0.036* | 0.043** | 0.014 | 0.021 |
| | (0.019) | (0.019) | (0.014) | (0.015) |
| Log#AbnMessages | 0.089 | 0.044 | 0.041 | -0.004 |
| | (0.060) | (0.072) | (0.054) | (0.066) |
| Observations | 1,007 | 822 | 1,007 | 822 |
| Adjusted R-Squared | 0.440 | 0.473 | 0.455 | 0.490 |

**Table 9**
**GIF Sentiment, GIF Disagreement, and Trading Volume**

This table reports the regression estimates of Equation (5), examining the relationship between sentiment and total trading volume. The analysis uses intraday data from September 2020 to December 2023 to construct return and sentiment measures at the 30-minute interval level. The dependent variable, LogTotalVol, is defined as the natural logarithm of one plus the total SPY trading volume accumulated within a given 30-minute interval t, as well as over forward-looking windows from $t$ + 30 minutes to $t$ + 1 day and from $t$ + 30 minutes to $t$ + 2 days. In Panel A, the main independent variable, |GIFsentiment|, is the absolute value of the appearance-weighted average sentiment of GIFs posted on Stocktwits during 30-minute interval $t$. In Panel B, we use GIFDisagreement, an intraday dispersion-based sentiment measure during interval $t$. GIFsentiment, GIFDisagreement, TEXTsentiment, SELFDEC, BW, ICS, and MEDIAsentiment are standardized to have zero mean and unit variance. Standard errors (reported in parentheses) are computed using a moving block bootstrap as described in the main text. *, **, and *** denote significance at the 10%, 5%, and 1% level, respectively. Variable definitions are in Table A2.

Panel A: GIF Sentiment and Total Trading Volume

| VARIABLES | (1) LogTotalVol(t) | (2) LogTotalVol [$t$ + 30m, $t$ + 1d] | (3) LogTotalVol(t) [$t$ + 30m, $t$ + 2d] | (4) LogTotalVol(t) | (5) LogTotalVol [$t$ + 30m, $t$ + 1d] | (6) LogTotalVol(t) [t + 30m, $t$ + 2d] |
|---|---|---|---|---|---|---|
| \|GIFsentiment\| | 0.160*** | 0.039* | 0.042** | 0.358*** | 0.068*** | 0.073*** |
| | (0.056) | (0.020) | (0.019) | (0.051) | (0.017) | (0.017) |
| \|TEXTsentiment\| | | | | -0.008 | 0.033** | 0.038*** |
| | | | | (0.084) | (0.016) | (0.015) |
| \|SELFDEC\| | | | | 0.035 | -0.084*** | -0.089*** |
| | | | | (0.052) | (0.018) | (0.018) |
| \|BW\| | | | | -0.175*** | -0.062* | -0.064* |
| | | | | (0.063) | (0.036) | (0.036) |
| \|ICS\| | | | | -0.172*** | -0.081** | -0.081** |
| | | | | (0.056) | (0.032) | (0.034) |
| \|MEDIAsentiment\| | | | | 0.009 | -0.012 | 0.000 |
| | | | | (0.052) | (0.021) | (0.019) |
| Controls | YES | YES | YES | YES | YES | YES |
| Observations | 10,493 | 10,493 | 10,493 | 10,493 | 10,395 | 10,493 |
| Adjusted R-squared | 0.065 | 0.241 | 0.261 | 0.193 | 0.400 | 0.411 |

Panel B: GIF Disagreement and Total Trading Volume

| VARIABLES | (1) LogTotalVol(t) | (2) LogTotalVol [$t$ + 30m, $t$ + 1d] | (3) LogTotalVol(t) [$t$ + 30m, $t$ + 2d] | (4) LogTotalVol(t) | (5) LogTotalVol [$t$ + 30m, $t$ + 1d] | (6) LogTotalVol(t) [$t$ + 30m, $t$ + 2d] |
|---|---|---|---|---|---|---|
| GIFDisagreement | 0.116*** | 0.070*** | 0.068*** | 0.102*** | 0.051*** | 0.050*** |
| | (0.013) | (0.007) | (0.006) | (0.015) | (0.006) | (0.006) |
| \|TEXTsentiment\| | | | | 0.108** | 0.032* | 0.038** |
| | | | | (0.044) | (0.019) | (0.018) |
| \|SELFDEC\| | | | | -0.043 | -0.086*** | -0.086*** |
| | | | | (0.045) | (0.018) | (0.018) |
| \|BW\| | | | | -0.083 | -0.092*** | -0.090** |
| | | | | (0.058) | (0.036) | (0.035) |
| \|ICS\| | | | | -0.059 | -0.101*** | -0.098*** |
| | | | | (0.050) | (0.033) | (0.034) |
| \|MEDIAsentiment\| | | | | -0.038 | -0.033*** | -0.030*** |
| | | | | (0.025) | (0.010) | (0.010) |
| Controls | YES | YES | YES | YES | YES | YES |
| Observations | 10,493 | 10,493 | 10,493 | 10,493 | 10,493 | 10,493 |
| Adjusted R-squared | 0.092 | 0.304 | 0.324 | 0.096 | 0.328 | 0.349 |

## Table 10
## Regression of Equity and Bond Fund Flows on the Sentiment Proxies

This table reports the regression estimates of Equation (6) from September 2020 to October 2024. In Panel A, the dependent variable, EFF is the daily net equity fund flow scaled by the fund's assets under management. In Panel B, BFF is the daily net bond fund flow scaled by the fund's assets under management. For both Panel A and B, we regress EFF and BFF on day-of-week and month-of-year dummies to remove seasonality and use the residuals as the dependent variables. Standard errors (reported in parentheses) are computed using a moving block bootstrap as described in the main text. *, **, and *** denote significance at the 10%, 5%, and 1% level, respectively. Variable definitions are in Table A2.

Panel A: Equity Fund Flow

| VARIABLES | (1)<br>EFF(*t*) | (2)<br>EFF[*t* +1, *t* + 5] | (3)<br>EFF(*t*) | (5)<br>EFF[t + 1, *t* + 5] |
|---|---|---|---|---|
| GIFsentiment | 0.004 | 0.005** | 0.001 | 0.007** |
| | (0.004) | (0.002) | (0.004) | (0.003) |
| TEXTsentiment | | | -0.003 | -0.005* |
| | | | (0.006) | (0.003) |
| SELFDEC | | | -0.012* | 0.006 |
| | | | (0.006) | (0.005) |
| BW | | | -0.004 | 0.003 |
| | | | (0.004) | (0.003) |
| ICS | | | -0.003 | -0.003 |
| | | | (0.005) | (0.004) |
| MEDIAsentiment | | | 0.016*** | -0.002 |
| | | | (0.005) | (0.002) |
| Controls | YES | YES | YES | YES |
| Observations | 1,007 | 822 | 1,007 | 822 |
| Adjusted R-squared | 0.038 | 0.085 | 0.085 | 0.159 |

Panel B1: Bond Fund Flow

| VARIABLES | (1)<br>BFF(t) | (2)<br>BFF[*t* + 1, *t* + 5] | (3)<br>BFF(t) | (5)<br>BFF[*t* + 1, *t* + 5] |
|---|---|---|---|---|
| GIF sentiment | -0.049* | -0.054*** | -0.006 | -0.031* |
| | (0.026) | (0.015) | (0.032) | (0.017) |
| Text sentiment | | | -0.003 | 0.083*** |
| | | | (0.033) | (0.030) |
| SELFDEC | | | -0.064 | -0.062* |
| | | | (0.042) | (0.032) |
| BW sentiment | | | 0.010 | 0.008 |
| | | | (0.029) | (0.020) |
| ICS | | | 0.030 | 0.022 |
| | | | (0.041) | (0.026) |
| Media sentiment | | | -0.068* | -0.040** |
| | | | (0.037) | (0.017) |
| Controls | YES | YES | YES | YES |
| Observations | 1,007 | 822 | 1,007 | 822 |
| Adjusted R-squared | 0.064 | 0.128 | 0.133 | 0.144 |

**Table A1**
**GIFs with Top 25 Highest and Top 25 Lowest Valence**

This table reports the GIFs with the top 25 highest and lowest sentiment throughout our sample period. Sentiment for each GIF is calculated as the difference between the total number of bullish declarations and bearish declarations, divided by the total number of appearances for each GIF during our sample period. GIFs on Giphy.com share a uniform URL structure: https://giphy.com/gifs/{giphy_id}. To view the animated GIFs, substitute {giphy_id} with the corresponding value in the Giphy_ID column.

| GIFs with lowest valence | | GIFs with highest valence | |
|---|---|---|---|
| Giphy_ID | Valence | Giphy_ID | Valence |
| l41YdlqVlryxP91Sw | -0.957 | xTiTnkt1IjaaTWoPny | 0.996 |
| WRbRNyAjA0mIw | -0.884 | 2cpPfXUit2JSU | 0.995 |
| kfd19XS70QrTQmsw7k | -0.866 | f9AxU1ieQdqfHbbf13 | 0.995 |
| UfX4XeBMXWmNoGvBVK | -0.816 | ibhRKzDTJ7O12 | 0.991 |
| IQ9KefLJHfJPq | -0.779 | XZVYAstOMLUDndgFPS | 0.972 |
| xA5oN4RDaCneQfcC8x | -0.761 | YpwSw00aOaoIhPVSAF | 0.971 |
| 9gGi02YPpLo2ueSxvh | -0.744 | PnahEQ7Ify1JvhQrag | 0.968 |
| NUZ5OqHdbknHa | -0.744 | mCsoBwnIyB5PSYwsAo | 0.943 |
| utMwbVuNZSSlvej2En | -0.734 | 9z8Jpk8Sl9QRrWqXlR | 0.940 |
| l0K4puBUN4w6G4ksE | -0.727 | 3ohjUQnfcfYR0QA3Yc | 0.940 |
| JpN6nbJqz5l3mbMnod | -0.722 | 1AeRujyfNSXi35GDU8 | 0.938 |
| 3LcOi1fXmCzNaYyemC | -0.691 | qrfjUqL8RPqmNB5LQM | 0.936 |
| m6tNZJt9cG3ss | -0.676 | ULoie47jnvxwtkx90t | 0.935 |
| dvZSDOywoCM4Sro65Q | -0.674 | QMyF0t2nkwNNt2sJp2 | 0.934 |
| ERIB4ws3cw17uWN4mF | -0.673 | dYyM2kbL89gYVS73IM | 0.933 |
| 9l84gf0TK6B7C3UCZp | -0.671 | 3o6wNKjI7XkipBHUjK | 0.928 |
| LkuPxRS0F6gmc | -0.661 | 9rjzS2QYAk1paKD7uk | 0.926 |
| 9detkWt4jBdhVm0UCk | -0.658 | HGvjR72DXRHWw | 0.925 |
| y31rRE5h3wyPXey8vx | -0.654 | HxtPXNp0ahLyM | 0.925 |
| C5ZIna5oroan9cdHz9 | -0.652 | VF6zQwFDlpE12FzBUB | 0.922 |
| VjzHEo2kXOxtJOJRCS | -0.649 | eu5jaVImGyKnsohsGy | 0.920 |
| JmUrefXyfqajgUNd0p | -0.642 | bEIQpE3d1sENO | 0.919 |
| 11Y9TiZzmEBe25QRSw | -0.640 | 6xE1FNcorRInS | 0.915 |
| LcS32DLbuturC | -0.632 | l4Ep9KQRRXtyjkIWQ | 0.914 |
| w4NAKAenurl8k | -0.623 | 3xz2BzSNxkwPqF8Wdy | 0.914 |

**Table A2**
**Variable Definitions**

| Variable | Definition | Source |
|---|---|---|
| ADS | U.S. macroeconomic activity index. | Aruoba, Diebold, and Scotti (2009) |
| Bond Fund Flow (BFF) | Daily aggregated mutual fund flow that specialize in US bonds. | EPFR |
| BW | Baker-Wugler monthly sentiment index. | https://pages.stern.nyu.edu/~jwurgler/ |
| Covid_Index | Daily index based on COVID-19's lockdown restrictions, including school closures, workplace closures, cancellations of public events, restrictions on gathering sizes, closures of public transport, stay-at-home requirements, restrictions on internal movement, and restrictions on international travel. The index ranges from 0 to 21. | University of Oxford's COVID-19 government response tracker |
| DCC | Daily average cloud cover using hourly values from 6am to 12pm across the country's weather stations. We deseasonalize the average daily cloud cover by subtracting each week's mean cloud cover from each daily mean following Hirshleifer and Shumway (2003). | National Oceanic and Atmospheric Administration |
| Equity Fund Flow (EFF) | Daily aggregated mutual fund flow that specialize in US equity. | EPFR |
| EPU | News-based measure of U.S. economic policy uncertainty. | Baker, Bloom, and Davis (2016) |
| GIFsentiment | Daily average sentiment of GIFs in all postings with cashtags (including both single and multiple cashtags), standardized to have a zero mean and unit variance. | Stocktwits |
| ICS | Monthly consumer confidence index. | University of Michigan's Surveys of Consumers |
| Log#AbnMessages | The deviation from the median number of messages posted over the prior 10 trading days. | Stocktwits |
| Log#EA | Daily natural logarithm of one plus the number of earnings announcements. | COMPUSTAT |
| MEDIAsentiment | Daily measure of traditional news media sentiment. | RavenPack |
| SELFDEC | Daily average sentiment of users' self-declarations in all non-GIF postings, standardized to have a zero mean and unit variance. | Stocktwits |
| SPX (%) | Daily return of S&P 500 Index. | CRSP |
| SPY (%) | 30-min interval intraday return of SPDR S&P 500 ETF. | TAQ |
| Textsentiment | Daily average sentiment of text in all postings, standardized to have a zero mean and unit variance. | Stocktwits |

**Table A3**
**User Characteristics and The Use of GIFs and Self-Declarations**

This table reports user characteristics associated with the use of GIFs and self-declared sentiment. Panel A regresses an indicator for GIF usage (GIF_Dummy) and the logarithm of the number of GIFs posted (Log(#GIF)) for each user *i* on indicators for self-declared trading experience (novice, intermediate, and professional). Panel B regresses an indicator for self-declared sentiment usage (SELFDEC_Dummy) and the logarithm of the number of self-declaration posts (Log(#SELFDEC)) on the same set of trading-experience indicators. Panel C reports Pearson correlations between GPT-4o-coded GIF sentiment and self-declaration-based GIF sentiment; see Section 2.5 for a detailed discussion. *, **, and *** denote significance at the 10%, 5%, and 1% level, respectively.

Panel A: User Characteristics and The Use of GIFs

| DEP.VAR= | $GIF_Dummy_i$ | $Log(\#GIF)_i$ |
|---|---|---|
| $\mathbb{1}Declared_Novice_i$ | 0.026*** | 0.023*** |
| | (0.004) | (0.008) |
| $\mathbb{1}Declared_Intermediate_i$ | -0.001 | -0.028*** |
| | (0.003) | (0.006) |
| $\mathbb{1}Declared_Professional_i$ | -0.039*** | -0.101*** |
| | (0.004) | (0.010) |
| $Activedays_i$ | 0.163*** | 0.450*** |
| | (0.000) | (0.001) |
| $Log(\#followers)_i$ | -0.026*** | -0.038*** |
| | (0.001) | (0.001) |
| $\mathbb{1}User_bio_i$ | 0.059*** | 0.159*** |
| | (0.002) | (0.005) |
| $Mean(Text\ sentiment)_i$ | 0.001 | 0.043*** |
| | (0.002) | (0.006) |
| $Mean(\#Word)_i$ | -0.002*** | -0.003*** |
| | (0.000) | (0.000) |
| $Mean(\#Number)_i$ | -0.007*** | -0.013*** |
| | (0.000) | (0.001) |
| $Mean(\#Emoji)_i$ | 0.015*** | 0.047*** |
| | (0.001) | (0.001) |
| | | |
| Observations | 474,488 | 474,488 |
| Adjusted R-squared | 0.237 | 0.325 |

Panel B: User Characteristics and The Use of Self-Declarations

| DEP.VAR= | $SELFDEC_Dummy_i$ | $Log(\#SELFDEC)_i$ |
|---|---|---|
| $\mathbb{1}$Declared Novice$_i$ | -0.008* | -0.081*** |
| | (0.005) | (0.011) |
| $\mathbb{1}$Declared Intermediate$_i$ | 0.020*** | 0.014* |
| | (0.003) | (0.008) |
| $\mathbb{1}$Declared Professional$_i$ | 0.050*** | 0.128*** |
| | (0.006) | (0.014) |
| Activedays$_i$ | -0.032*** | 0.275*** |
| | (0.001) | (0.002) |
| Log(#followers)$_i$ | -0.000 | 0.013*** |
| | (0.001) | (0.002) |
| $\mathbb{1}$User_bio$_i$ | -0.036*** | -0.120*** |
| | (0.003) | (0.007) |
| Mean(Text sentiment)$_i$ | 0.063*** | 0.090*** |
| | (0.003) | (0.008) |
| Mean(#Word)$_i$ | 0.001*** | -0.000 |
| | (0.000) | (0.000) |
| Mean(#Number)$_i$ | 0.006*** | 0.012*** |
| | (0.000) | (0.001) |
| Mean(#Emoji)$_i$ | -0.003*** | -0.004** |
| | (0.001) | (0.002) |
| Observations | 474,488 | 474,488 |
| Adjusted R-squared | 0.011 | 0.090 |

Panel C: Correlation Between GIF Sentiment from Self-declarations and GPT-Coded Sentiment

| VARIABLES | (1)<br>ST-GIFsentiment$_{continuous}$ | (2)<br>ST-GIFsentiment$_{binary}$ |
|---|---|---|
| Gpt4o-GIFsentiment$_{continuous}$ | 0.306*** | |
| | <0.0001 | |
| Gpt4o-GIFsentiment$_{binary}$ | | 0.537*** |
| | | <0.0001 |

**Table A4**
**Robustness to Winsorize Returns at the top and bottom 1%, 5%, and 10% levels.**

This table reports the regression estimates of Equation (3). The dependent variable is the Standard and Poor's 500 Index (SPX) return at alternative windows. We Winsorize returns at the top and bottom 1%, 5%, and 10% levels. We multiply returns by 100 to interpret coefficients as percentage points. The main independent variable, GIFsentiment is the daily appearance-weighted average sentiment of GIFs posted on Stocktwits. Sentiment measures are standardized to have zero mean and unit variance. Standard errors (reported in parentheses) are computed using a moving block bootstrap as described in the main text. *, **, and *** denote significance at the 10%, 5%, and 1% level, respectively. Variable definitions are in Table A2.

Panel A: GIF Sentiment Alone

| | (1) | (2) | (3) | (4) | (5) | (6) | (7) | (8) | (9) |
|---|---|---|---|---|---|---|---|---|---|
| | Winsorize at top and bottom 1 percentile | | | Winsorize at top and bottom 5 percentile | | | Winsorize at top and bottom 10 percentile | | |
| VARIABLES | Ret($t$) | Ret[$t$ + 1, $t$ + 5] | Ret[$t$ + 1, $t$ + 20] | Ret($t$) | Ret[$t$ + 1, $t$ + 5] | Ret[$t$ + 1, $t$ + 20] | Ret($t$) | Ret[$t$ + 1, $t$ + 5] | Ret[$t$ + 1, $t$ + 20] |
| GIFsentiment | 0.273*** | 0.057 | -0.915*** | 0.251*** | 0.053 | -0.709** | 0.228*** | 0.061 | -0.555** |
| | (0.065) | (0.158) | (0.323) | (0.058) | (0.140) | (0.295) | (0.051) | (0.124) | (0.260) |
| EPU | 0.001 | 0.001 | 0.004 | 0.001 | 0.001 | 0.004 | 0.001 | 0.000 | 0.004 |
| | (0.001) | (0.002) | (0.003) | (0.001) | (0.002) | (0.003) | (0.001) | (0.001) | (0.002) |
| ADS | -0.019 | 0.432 | 1.749** | -0.014 | 0.407 | 1.538** | -0.015 | 0.360 | 1.278** |
| | (0.110) | (0.309) | (0.694) | (0.098) | (0.256) | (0.613) | (0.086) | (0.227) | (0.564) |
| Ret($t$) | | -0.147 | -0.339* | | -0.166 | -0.373* | | -0.174 | -0.379* |
| | | (0.143) | (0.204) | | (0.141) | (0.198) | | (0.143) | (0.197) |
| Ret[$t$ - 5, $t$ - 1] | -0.052** | -0.089 | -0.319** | -0.055** | -0.087 | -0.327** | -0.057** | -0.094 | -0.302** |
| | (0.026) | (0.065) | (0.143) | (0.027) | (0.061) | (0.144) | (0.027) | (0.059) | (0.146) |
| Ret[$t$ - 21, $t$ - 6] | 0.000 | -0.065 | -0.214* | -0.003 | -0.068 | -0.168 | -0.002 | -0.064 | -0.106 |
| | (0.016) | (0.044) | (0.116) | (0.015) | (0.044) | (0.109) | (0.015) | (0.044) | (0.113) |
| Log#EA | -0.006 | 0.013 | -0.154 | -0.006 | 0.020 | -0.101 | -0.007 | 0.014 | -0.078 |
| | (0.043) | (0.113) | (0.233) | (0.036) | (0.097) | (0.220) | (0.032) | (0.085) | (0.195) |
| Log#AbnMessages | -0.013 | 0.047 | 0.130 | 0.006 | 0.029 | 0.107 | 0.012 | 0.026 | 0.053 |
| | (0.131) | (0.409) | (0.786) | (0.118) | (0.380) | (0.690) | (0.107) | (0.355) | (0.623) |
| Observations | 1,007 | 1,002 | 988 | 1,007 | 1,002 | 988 | 1,007 | 1,002 | 988 |
| Adjusted R-squared | 0.051 | 0.026 | 0.166 | 0.055 | 0.027 | 0.148 | 0.058 | 0.029 | 0.130 |

Panel B: Six Sentiment Measures

| | (1) | (2) | (3) | (4) | (5) | (6) | (7) | (8) | (9) |
|---|---|---|---|---|---|---|---|---|---|
| | Winsorize at top and bottom 1 percentile | | | Winsorize at top and bottom 5 percentile | | | Winsorize at top and bottom 10 percentile | | |
| VARIABLES | Ret(*t*) | Ret[*t* + 1, *t* + 5] | Ret[*t* + 1, *t* + 20] | Ret(*t*) | Ret[*t* + 1, *t* + 5] | Ret[*t* + 1, *t* + 20] | Ret(*t*) | Ret[*t* + 1, *t* + 5] | Ret[*t* + 1, *t* + 20] |
| GIFsentiment | 0.287** | -0.007 | -1.647*** | 0.237** | -0.016 | -1.409*** | 0.200** | 0.003 | -1.181** |
| | (0.126) | (0.279) | (0.638) | (0.111) | (0.241) | (0.539) | (0.093) | (0.211) | (0.479) |
| TEXTsentiment | 0.155** | -0.211 | 0.212 | 0.144** | -0.129 | 0.189 | 0.128** | -0.059 | 0.223 |
| | (0.071) | (0.183) | (0.424) | (0.064) | (0.164) | (0.395) | (0.056) | (0.146) | (0.363) |
| SELFDEC | 0.014 | 0.340 | 1.890*** | 0.046 | 0.337 | 1.778*** | 0.056 | 0.285 | 1.491*** |
| | (0.149) | (0.309) | (0.711) | (0.140) | (0.270) | (0.578) | (0.121) | (0.232) | (0.527) |
| BW | -0.345*** | -0.096 | -1.237** | -0.314*** | -0.166 | -1.089*** | -0.272*** | -0.177 | -0.829** |
| | (0.089) | (0.235) | (0.500) | (0.072) | (0.208) | (0.413) | (0.060) | (0.177) | (0.364) |
| ICS | -0.282*** | -0.064 | 0.042 | -0.240*** | -0.037 | 0.125 | -0.203*** | -0.014 | 0.180 |
| | (0.088) | (0.292) | (0.718) | (0.080) | (0.236) | (0.557) | (0.071) | (0.198) | (0.469) |
| MEDIAsentiment | 0.537*** | -0.181 | -0.569 | 0.462*** | -0.081 | -0.468 | 0.392*** | -0.011 | -0.440 |
| | (0.114) | (0.196) | (0.372) | (0.090) | (0.173) | (0.332) | (0.074) | (0.149) | (0.295) |
| EPU | 0.001* | 0.002 | 0.006** | 0.001 | 0.001 | 0.005** | 0.001 | 0.001 | 0.006** |
| | (0.001) | (0.002) | (0.003) | (0.001) | (0.002) | (0.003) | (0.001) | (0.002) | (0.002) |
| ADS | 0.052 | 0.685* | 2.374*** | 0.049 | 0.630** | 2.111*** | 0.034 | 0.525* | 1.699*** |
| | (0.117) | (0.371) | (0.676) | (0.107) | (0.305) | (0.580) | (0.096) | (0.271) | (0.542) |
| Ret(*t*) | | -0.110 | -0.453** | | -0.175 | -0.552** | | -0.230 | -0.578*** |
| | | (0.149) | (0.221) | | (0.148) | (0.221) | | (0.148) | (0.223) |
| Ret[*t* - 5, *t* - 1] | -0.060** | -0.108 | -0.446*** | -0.062** | -0.120* | -0.500*** | -0.065** | -0.141** | -0.496*** |
| | (0.028) | (0.073) | (0.130) | (0.028) | (0.068) | (0.134) | (0.028) | (0.067) | (0.141) |
| Ret[*t* - 21, *t* - 6] | -0.026 | -0.107** | -0.449*** | -0.033* | -0.131*** | -0.444*** | -0.031* | -0.135*** | -0.374*** |
| | (0.017) | (0.044) | (0.126) | (0.017) | (0.042) | (0.118) | (0.018) | (0.042) | (0.125) |
| Log#EA | 0.072 | 0.106 | 0.047 | 0.064 | 0.120 | 0.099 | 0.053 | 0.102 | 0.077 |
| | (0.053) | (0.125) | (0.211) | (0.046) | (0.105) | (0.200) | (0.040) | (0.091) | (0.182) |
| Log#AbnMessages | -0.109 | -0.079 | 0.138 | -0.065 | -0.071 | 0.176 | -0.041 | -0.049 | 0.193 |
| | (0.167) | (0.484) | (0.973) | (0.151) | (0.442) | (0.855) | (0.135) | (0.408) | (0.769) |
| Observations | 822 | 822 | 822 | 822 | 822 | 822 | 822 | 822 | 822 |
| Adjusted R-squared | 0.226 | 0.053 | 0.319 | 0.234 | 0.057 | 0.325 | 0.230 | 0.060 | 0.305 |

**Table A5**
**Robustness to GIF Appearance Frequency Thresholds**

This table reports the regression estimates of Equation (3). The dependent variable is the Standard and Poor's 500 Index (SPX) return at alternative windows. We multiply returns by 100 to interpret coefficients as percentage points. The main independent variable, GIFsentiment is the daily appearance-weighted average sentiment of GIFs posted on Stocktwits, using only GIFs whose cumulative appearances exceed the 50th or 75th percentile among all GIFs as of day *t*. Sentiment measures are standardized to have zero mean and unit variance. Standard errors (reported in parentheses) are computed using a moving block bootstrap as described in the main text. *, **, and *** denote significance at the 10%, 5%, and 1% level, respectively. Variable definitions are in Table A2.

Panel A: GIF Sentiment Alone

| | (1) | (2) | (3) | (4) | (5) | (6) |
|---|---|---|---|---|---|---|
| | Keep only GIFs whose cumulative appearances exceed the 50$^{th}$ percentile | | | Keep only GIFs whose cumulative appearances exceed the 75$^{th}$ percentile | | |
| VARIABLES | Ret(*t*) | Ret[*t* + 1, t + 5] | Ret[*t* + 1, *t* + 20] | Ret(*t*) | Ret[*t* + 1, *t* + 5] | Ret[*t* + 1, *t* + 20] |
| GIFsentiment | 0.271*** | 0.042 | -1.357*** | 0.252*** | 0.050 | -1.190*** |
| | (0.082) | (0.172) | (0.357) | (0.081) | (0.163) | (0.336) |
| EPU | 0.001* | 0.002 | 0.003 | 0.002* | 0.002 | 0.002 |
| | (0.001) | (0.002) | (0.003) | (0.001) | (0.002) | (0.003) |
| ADS | 0.023 | 0.387 | 1.500** | 0.062 | 0.394 | 1.330* |
| | (0.113) | (0.326) | (0.721) | (0.114) | (0.329) | (0.732) |
| Ret(*t*) | | -0.149 | -0.242 | | -0.150 | -0.269 |
| | | (0.147) | (0.210) | | (0.147) | (0.210) |
| Ret[*t* - 5, *t* - 1] | -0.046* | -0.071 | -0.241 | -0.043 | -0.071 | -0.262* |
| | (0.026) | (0.065) | (0.150) | (0.026) | (0.065) | (0.153) |
| Ret[*t* - 21, *t* - 6] | 0.011 | -0.049 | -0.214* | 0.011 | -0.049 | -0.215* |
| | (0.016) | (0.044) | (0.124) | (0.016) | (0.044) | (0.125) |
| Log#EA | -0.010 | 0.016 | -0.107 | -0.009 | 0.016 | -0.116 |
| | (0.045) | (0.120) | (0.243) | (0.045) | (0.121) | (0.249) |
| Log#AbnMessages | -0.029 | 0.030 | 0.196 | -0.025 | 0.028 | 0.188 |
| | (0.137) | (0.420) | (0.865) | (0.138) | (0.426) | (0.881) |
| Observations | 1,007 | 1,002 | 988 | 1,007 | 1,002 | 988 |
| Adjusted R-squared | 0.040 | 0.020 | 0.160 | 0.037 | 0.020 | 0.148 |

Panel B: Six Sentiment Measures

| | (1) | (2) | (3) | (4) | (5) | (6) |
|---|---|---|---|---|---|---|
| | Keep only GIFs whose cumulative appearances exceed the 50th percentile | | | Keep only GIFs whose cumulative appearances exceed the 75th percentile | | |
| VARIABLES | Ret(*t*) | Ret[*t* + 1, *t* + 5] | Ret[*t* + 1, *t* + 20] | Ret(*t*) | Ret[*t* + 1, *t* + 5] | Ret[*t* + 1, *t* + 20] |
| GIFsentiment | 0.242** | 0.017 | -1.547** | 0.224** | 0.012 | -1.531** |
| | (0.112) | (0.271) | (0.656) | (0.100) | (0.272) | (0.615) |
| TEXTsentiment | 0.131* | -0.252 | 0.301 | 0.124* | -0.251 | 0.376 |
| | (0.073) | (0.197) | (0.454) | (0.071) | (0.200) | (0.447) |
| SELFDEC | 0.101 | 0.236 | 1.120 | 0.125 | 0.239 | 1.004 |
| | (0.123) | (0.288) | (0.704) | (0.117) | (0.280) | (0.666) |
| BW | -0.335*** | -0.033 | -1.071* | -0.315*** | -0.031 | -1.190** |
| | (0.092) | (0.253) | (0.565) | (0.088) | (0.249) | (0.543) |
| ICS | -0.260*** | -0.009 | -0.093 | -0.239** | -0.009 | -0.272 |
| | (0.094) | (0.315) | (0.824) | (0.097) | (0.326) | (0.854) |
| MEDIAsentiment | 0.545*** | -0.216 | -0.673* | 0.542*** | -0.217 | -0.672* |
| | (0.118) | (0.203) | (0.403) | (0.117) | (0.201) | (0.401) |
| EPU | 0.001* | 0.002 | 0.006* | 0.001* | 0.002 | 0.006* |
| | (0.001) | (0.002) | (0.003) | (0.001) | (0.002) | (0.003) |
| ADS | 0.033 | 0.605 | 2.168*** | 0.032 | 0.605 | 2.179*** |
| | (0.124) | (0.398) | (0.738) | (0.124) | (0.401) | (0.735) |
| Ret(*t*) | | -0.093 | -0.306 | | -0.093 | -0.308 |
| | | (0.157) | (0.236) | | (0.158) | (0.235) |
| Ret[*t* - 5, *t* - 1] | -0.055** | -0.081 | -0.331** | -0.055** | -0.080 | -0.329** |
| | (0.028) | (0.074) | (0.145) | (0.028) | (0.074) | (0.145) |
| Ret[*t* - 21, *t* - 6] | -0.018 | -0.077* | -0.362*** | -0.018 | -0.077* | -0.359*** |
| | (0.019) | (0.046) | (0.135) | (0.018) | (0.046) | (0.134) |
| Log#EA | 0.067 | 0.090 | 0.038 | 0.065 | 0.090 | 0.052 |
| | (0.056) | (0.132) | (0.229) | (0.056) | (0.133) | (0.225) |
| Log#AbnMessages | -0.130 | -0.094 | 0.197 | -0.129 | -0.093 | 0.216 |
| | (0.177) | (0.488) | (1.031) | (0.179) | (0.488) | (1.021) |
| Observations | 822 | 822 | 822 | 822 | 822 | 822 |
| Adjusted R-squared | 0.207 | 0.042 | 0.262 | 0.205 | 0.042 | 0.264 |

**Table A6**
**Robustness to Post Estimation Check**

This table reports the regression estimates of Equation (3). We exclude observations with high influence as identified by the DFBETA test, where influence is defined as having an absolute DFBETA greater than $2/\sqrt{n}$, with n denoting the sample size. DFBETA quantifies the impact of each observation on the estimated regression coefficient for GIFsentiment by comparing the coefficient with and without that observation. The dependent variable is the Standard and Poor's 500 Index (SPX) return at alternative windows. We multiply returns by 100 to interpret coefficients as percentage points. The main independent variable, GIFsentiment, is our baseline measure that includes only GIFs with at least five declarations. Sentiment measures are standardized to have zero mean and unit variance. Standard errors (reported in parentheses) are computed using a moving block bootstrap as described in the main text. *, **, and *** denote significance at the 10%, 5%, and 1% level, respectively. Variable definitions are in Table A2.

Panel A: GIF Sentiment Alone

| VARIABLES | (1)<br>Ret(*t*)<br>Exclude 34 days with large DFBETA values | (2)<br>Ret[*t* + 1, *t* + 5]<br>Exclude 33 days with large DFBETA values | (3)<br>Ret[*t* + 1, *t* + 20]<br>Exclude 25 days with large DFBETA values |
|---|---|---|---|
| GIFsentiment | 0.195*** | -0.437*** | -1.394*** |
| | (0.060) | (0.136) | (0.323) |
| EPU | 0.000 | 0.001 | 0.003 |
| | (0.001) | (0.002) | (0.003) |
| ADS | 0.001 | 0.490* | 1.517** |
| | (0.092) | (0.253) | (0.737) |
| Ret(*t*) | | -0.077 | -0.117 |
| | | (0.139) | (0.223) |
| Ret[*t* - 5, *t* - 1] | -0.042* | -0.043 | -0.166 |
| | (0.023) | (0.058) | (0.147) |
| Ret[*t* - 21, *t* - 6] | -0.020 | -0.063 | -0.316** |
| | (0.014) | (0.042) | (0.123) |
| Log#EA | 0.036 | 0.087 | 0.104 |
| | (0.041) | (0.100) | (0.228) |
| Log#AbnMessages | -0.075 | 0.363 | 0.820 |
| | (0.117) | (0.416) | (0.726) |
| Observations | 973 | 969 | 963 |
| Adjusted R-squared | 0.034 | 0.064 | 0.260 |

Panel B: Six Sentiment Measures

| VARIABLES | (1)<br>Ret(*t*)<br>Exclude 52 days with large DFBETA values | (2)<br>Ret[*t* + 1, *t* + 5]<br>Exclude 49 days with large DFBETA values | (3)<br>Ret[*t* + 1, *t* + 20]<br>Exclude 54 days with large DFBETA values |
|---|---|---|---|
| GIFsentiment | 0.112* | 0.018 | -0.961** |
| | (0.066) | (0.183) | (0.419) |
| TEXTsentiment | 0.130** | -0.200 | 0.331 |
| | (0.065) | (0.228) | (0.321) |
| SELFDEC | 0.242*** | -0.127 | 0.114 |
| | (0.086) | (0.276) | (0.592) |
| BW | -0.451*** | -0.299 | -0.993* |
| | (0.071) | (0.238) | (0.544) |
| ICS | -0.301*** | 0.078 | 0.195 |
| | (0.074) | (0.334) | (0.776) |
| MEDIAsentiment | 0.514*** | -0.097 | -0.935** |
| | (0.075) | (0.169) | (0.376) |
| EPU | -0.001 | 0.001 | 0.004 |
| | (0.001) | (0.002) | (0.003) |
| ADS | 0.051 | 0.606 | 1.855*** |
| | (0.108) | (0.377) | (0.680) |
| Ret(*t*) | | -0.074 | -0.087 |
| | | (0.121) | (0.240) |
| Ret[*t* - 5, *t* - 1] | -0.044* | -0.126* | -0.325** |
| | (0.024) | (0.076) | (0.156) |
| Ret[*t* - 21, *t* - 6] | -0.047*** | -0.110** | -0.343*** |
| | (0.016) | (0.046) | (0.110) |
| Log#EA | 0.119*** | 0.212 | 0.209 |
| | (0.037) | (0.134) | (0.256) |
| Log#AbnMessages | -0.234 | 0.156 | 1.070 |
| | (0.144) | (0.394) | (0.884) |
| Observations | 770 | 773 | 768 |
| Adjusted R-squared | 0.244 | 0.107 | 0.304 |

**Table A7**
**Robustness to Excluding the Initial Adoption Period**

This table reports the regression estimates of Equation (3). We exclude observations from the initial adoption period of the GIF-sending function on Stocktwits.com, spanning September 1, 2020 to December 31, 2020. The dependent variable is the Standard and Poor's 500 Index (SPX) return at alternative windows. We multiply returns by 100 to interpret coefficients as percentage points. The main independent variable, GIFsentiment is the daily appearance-weighted average sentiment of GIFs posted on Stocktwits. Sentiment measures are standardized to have zero mean and unit variance. Standard errors (reported in parentheses) are computed using a moving block bootstrap as described in the main text. *, **, and *** denote significance at the 10%, 5%, and 1% level, respectively. Variable definitions are in Table A2.

Panel A: GIF Sentiment Alone

| VARIABLES | (1)<br>Ret(*t*) | (2)<br>Ret[*t* + 1, *t* + 5] | (3)<br>Ret[*t* + 1, *t* + 20] |
|---|---|---|---|
| GIFsentiment | 0.225*** | -0.377 | -1.788*** |
| | (0.056) | (0.250) | (0.583) |
| EPU | 0.000 | 0.001 | 0.004 |
| | (0.001) | (0.002) | (0.003) |
| ADS | -0.064 | 0.998*** | 1.439** |
| | (0.087) | (0.327) | (0.724) |
| Ret(*t*) | | -0.038 | 0.443* |
| | | (0.133) | (0.252) |
| Ret[*t* - 5, *t* - 1] | -0.041 | 0.031 | 0.208 |
| | (0.026) | (0.064) | (0.144) |
| Ret[*t* - 21, *t* - 6] | -0.009 | -0.099* | -0.316** |
| | (0.016) | (0.051) | (0.141) |
| Log#EA | 0.046 | 0.100 | -0.516** |
| | (0.030) | (0.116) | (0.213) |
| Log#AbnMessages | -0.072 | 0.500 | 1.222 |
| | (0.127) | (0.462) | (0.933) |
| Observations | 926 | 921 | 907 |
| Adjusted R-squared | 0.052 | 0.067 | 0.157 |

Panel B: Six Sentiment Measures

| VARIABLES | (1)<br>Ret(*t*) | (2)<br>Ret[*t* + 1, *t* + 5] | (3)<br>Ret[*t* + 1, *t* + 20] |
|---|---|---|---|
| GIFsentiment | 0.271** | -0.486 | -2.392*** |
| | (0.127) | (0.409) | (0.793) |
| TEXTsentiment | 0.114 | 0.031 | 0.911 |
| | (0.088) | (0.268) | (0.581) |
| SELFDEC | 0.108 | 0.437 | 1.345 |
| | (0.113) | (0.387) | (0.842) |
| BW | -0.316*** | -0.906*** | -3.107*** |
| | (0.091) | (0.262) | (0.593) |
| ICS | -0.150* | -0.296 | -0.792 |
| | (0.086) | (0.344) | (0.922) |
| MEDIAsentiment | 0.444*** | -0.009 | 0.362 |
| | (0.080) | (0.235) | (0.377) |
| EPU | 0.000 | 0.003* | 0.008** |
| | (0.001) | (0.002) | (0.004) |
| ADS | -0.068 | 1.588*** | 2.653*** |
| | (0.125) | (0.483) | (0.970) |
| Ret(*t*) | | -0.208 | -0.185 |
| | | (0.148) | (0.247) |
| Ret[*t* - 5, *t* - 1] | -0.040 | -0.069 | -0.116 |
| | (0.031) | (0.076) | (0.157) |
| Ret[*t* - 21, *t* - 6] | -0.011 | -0.159*** | -0.513*** |
| | (0.020) | (0.052) | (0.137) |
| Log#EA | 0.088* | 0.227 | -0.200 |
| | (0.049) | (0.142) | (0.237) |
| Log#AbnMessages | -0.239 | 0.465 | 1.271 |
| | (0.174) | (0.559) | (0.963) |
| Observations | 741 | 741 | 741 |
| Adjusted R-squared | 0.159 | 0.147 | 0.379 |